\documentclass[pdftex,twocolumn,epjc3]{svjour3}          % twocolumn

\RequirePackage[T1]{fontenc}

\smartqed  % flush right qed marks, e.g. at end of proof

\RequirePackage{graphicx}
\RequirePackage{mathptmx}      % use Times fonts if available on your TeX system
\RequirePackage{flushend}
\RequirePackage[numbers,sort&compress]{natbib}
\RequirePackage[colorlinks,citecolor=blue,urlcolor=blue,linkcolor=blue]{hyperref}
\usepackage{amssymb,amsmath,array}
\RequirePackage{xspace}
\usepackage{hyperref}
\def\jpsi {\ensuremath{{J\mskip -3mu/\mskip -2mu\psi\mskip 2mu}}\xspace}
\mathchardef\Upsilon="7107
\def\Y#1S{\ensuremath{\Upsilon{(#1S)}}\xspace}% no space before {...}!

\def\Dbar    {\kern 0.2em\overline{\kern -0.2em D}{}\xspace}

\def\Bbar    {\kern 0.18em\overline{\kern -0.18em B}{}\xspace}

\usepackage{relsize}
\def\babar{\mbox{\slshape B\kern-0.1em{\smaller A}\kern-0.1em
    B\kern-0.1em{\smaller A\kern-0.2em R}}}

\def\ccbar {\ensuremath{c\overline c}\xspace}
\def\bbbar {\ensuremath{b\overline b}\xspace}

\def\epem {\ensuremath{e^+e^-}\xspace}

\newcommand{\kev}{\ensuremath{\mathrm{\,ke\kern -0.1em V}}\xspace}
\newcommand{\mev}{\ensuremath{\mathrm{\,Me\kern -0.1em V}}\xspace}
\newcommand{\mevcc}{\ensuremath{{\mathrm{\,Me\kern -0.1em V\!/}c^2}}\xspace}
\newcommand{\gev}{\ensuremath{\mathrm{\,Ge\kern -0.1em V}}\xspace}
\newcommand{\gevcc}{\ensuremath{{\mathrm{\,Ge\kern -0.1em V\!/}c^2}}\xspace}
\newcommand{\tev}{\ensuremath{\mathrm{\,Te\kern -0.1em V}}\xspace}
\newcommand{\tevcc}{\ensuremath{{\mathrm{\,Te\kern -0.1em V\!/}c^2}}\xspace}
\newcommand{\ev}{\ensuremath{\mathrm{\,e\kern -0.1em V}}\xspace}

\def\mum  {\ensuremath{{\,\mu\rm m}}\xspace}%% mu meter

\def\nb   {\ensuremath{{\rm \,nb}}\xspace}

\def\cms  {\ensuremath{{\rm \,cm}^{-2} {\rm s}^{-1}}\xspace}

\newcommand{\EPEM}{\ensuremath{e^+e^-}\xspace}

\newcommand{\GG}{\ensuremath{\gamma\gamma}\xspace}
\newcommand{\GE}{\ensuremath{\gamma e}\xspace}
\newcommand{\LGG}{\ensuremath{\mathcal{L}_{\gamma\gamma}}}

\newcommand{\LEE}{\ensuremath{\mathcal{L}_{ee}}\xspace}
\newcommand{\LEPEM}{\ensuremath{\mathcal{L}_{e^+e^-}}\xspace}
\newcommand{\WGG}{\ensuremath{W_{\gamma\gamma}}\xspace}
\newcommand{\GGG}{\ensuremath{\Gamma_{\gamma\gamma}}\xspace}

\newcommand{\be}{\begin{equation}}
\newcommand{\ee}{\end{equation}}
\newcommand{\bc}{\begin{center}}
\newcommand{\ec}{\end{center}}
\newcommand{\bi}{\begin{itemize}}
\newcommand{\ei}{\end{itemize}}
\newcommand{\ben}{\begin{enumerate}}
\newcommand{\een}{\end{enumerate}}

\journalname{Eur. Phys. J. C}

\begin{document}

\title{\boldmath Opportunities for studying $\mathit C$-even resonances \\ at a 3--12 GeV photon collider
%\thanksref{t1}
}


\author{K.~I.~Beloborodov\thanksref{addr1,addr2}
        \and
        T.~A.~Kharlamova\thanksref{addr1,addr2}
        \and
        G.~Moortgat-Pick\thanksref{addr3}
        \and
        V.~I.~Telnov~\thanksref{addr1,addr2,c1}
}

%\thankstext[$\star$]{t1}{Thanks to the title}
\thankstext[$\star$]{c1}{Corresponding author, e-mail:telnov@inp.nsk.su}
%\thankstext{e1}{e-mail: magic1@xxx.xx}
%\thankstext{e2}{e-mail: magic2@xxx.xx}
%\thankstext{e3}{e-mail: telnov@inp.nsk.su}

\institute{Budker Institute of Nuclear Physics, 630090, Novosibirsk, Russia\label{addr1}
          \and
          Novosibirsk State University, 630090, Novosibirsk, Russia\label{addr2}
          \and
          II. Inst. f. Theoret. Physics, University of Hamburg,  22761, Hamburg, Germany\label{addr3}
%          \and
%          \emph{Present Address:} Street, City, Country\label{addr3}
}

\date{Received: date / Accepted: date}

\maketitle

\begin{abstract}

Recently, a \GG\ collider based on the existing 17.5 GeV  linac of the European XFEL has been proposed. High-energy photons will be generated by Compton scattering of laser photons with a wavelength of 0.5--1 \mum on electrons. Such a photon collider covers the range of invariant masses $W_{\gamma\gamma} <12$ \gevcc. The physics program includes spectroscopy of $\mathit C$-even resonances ($c$-, $b$-quarkonia, 4-quark states, glueballs) in various $J^P$ states. Variable circular and linear polarizations will help in determining the quantum numbers. In this paper, we present a summary of measured and predicted two-photon widths of various resonances in the mass region 3--12 \gevcc  and investigate the experimental possibility of observing the\-se heavy two-photon resonances under the conditions of a large multi-hadron background. Registration of all final particles is assumed. The minimum values of $\Gamma_{\GG}(W)$ are obtained at which resonances can be detected at a $5\sigma$ confidence level in one year of operation .

\end{abstract}

\section{Introduction} \vspace{-0.2cm}
Gamma-gamma collisions have been studied  since the 1970s at \epem storage rings in collisions of virtual photons ($\gamma^*$). Two-photon physics is complementary to the \EPEM physics program: in \epem collisions, $C$-odd resonances with $J^P=1^-$ are produced, while in \GG\ collisions $C$-even resonances with various spins $J\neq 1$ are produced.  The first such resonance  ($\eta^{\prime}$) was observed in 1979 with detector Mark I at SPEAR~\cite{EtaP79}, followed by many two-photon results from all \epem\ facilities. Many results have been obtained at the high-luminosity KEKB and PEP-II~\cite{Bphys}, and studies continue at SuperKEKB. The number of virtual photons per electron is rather small, therefore \LGG $\ll$ \LEPEM (however, it is complementary).

The future of the \GG\ physics is closely connected with photon colliders based on high-energy linear colliders.  At linear colliders, beams are used only once, which makes possible $e \to \gamma$ conversion by Compton backscattering of laser light just before the interaction point, thus obtaining  \GG, \GE\  beams with a luminosity comparable to that in \EPEM\ collisions~\cite{GKST81r,GKST81e,GKST83,GKST84}.
Since the late 1980s, \GG\ colliders have been considered a natural part of all linear collider projects; conceptual~\cite{NLC,TESLAcdr,JLC} and pre-technical designs~\cite{TESLATDR,telnov-acta2} have been published. The photon collider is being considered as one of the Higgs factory options~\cite{Asner,Telnov-higgs,Telnov-higgs-pc}.  However, no linear collider has yet been approved, and the future is rather unclear. Recently, V.~Telnov proposed a photon collider~\cite{telnov-gg12} based on the electron linac of the existing linac of the European XFEL~\cite{xfel}. By pairing its 17.5 \gev electron beam with a 0.5 \mum laser, one can obtain a photon collider with a center-of-mass energy $W_{\GG} \leq 12$ \gevcc.
While the region $W_{\GG} <$ 4--5 \gevcc can be studied at SuperKEKB, in the region $W_{\GG}=$ 5--12 \gevcc the photon collider will have no competition in the study of a large number of \bbbar resonances, tetraquarks, or mesonic molecules.

  In this paper, we investigate the question of the very possibility of observing and studying heavy $C$-even resonances in the presence of a large hadronic background. The effective cross section of resonance production is proportional to $\Gamma_{\GG}/M_R^2$;  for bottomonium (\bbbar) states, this value is two orders of magnitude smaller than for charmonium (\ccbar) states.  At the same time, the cross section of the background  $\GG\ \to hadrons$ process in this energy region is nearly constant. At these "intermediate" energies, the angular distribution of hadronic backgrounds still differs not much from the isotropic distribution in resonance decays (for $J=0$), so the possibility of suppressing the background was not at all obvious.

  In Sect.~2, we summarize theoretical predictions on \GG widths $\Gamma_{\GG}$ of resonances in this energy region and give formulas for production cross sections in \GG\ collisions. In Sect.~3, main parameters of the \GG\ collider are presented and the differential luminosity $d\LGG/d\WGG$ is compared with that at SuperKEKB. In Sect.~4, we consider methods to suppress hadronic backgrounds (using realistic simulation) and determine detection efficiencies after background suppression. Finally, we find the values of $\GGG(W)$ for which  resonances can be observed at a $5\sigma$ confidence level in one year of the \GG collider operation.
\section{Two-photon processes: general features}
%
%\subsection{General features of luminosities and cross sections at \GG\ colliders}
The spectrum of photons after Compton backscattering is broad, with a
characteristic peak at maximum energies. Photons can have circular or
linear polarizations depending on their energies and polarizations of
initial electrons and laser photons. Due to the angle-energy correlation,
in Compton scattering the \GG\ luminosity cannot be described a by
convolution of photon spectra.  Due to complexity of processes in
the conversion and interaction regions, the accuracy of prediction by
simulation will be rather poor; therefore, one should measure all
luminosity properties experimentally using well known QED processes~\cite{Pak}.

Generally, the number of events in \GG\ collision is given
by~\cite{GKST84,Pak}
\begin{equation}
d\dot{N}_{\GG\ \to X} = dL_{\GG} \sum_{i,j=0}^3 \langle \xi_i \tilde{\xi_j}
\rangle \sigma_{ij},
\vspace*{-0.1cm}
\label{dn1}
\end{equation}
where $\xi_i$ are Stokes parameters, $\xi_2\equiv \lambda_{\gamma}$ is
the circular polarization, $\sqrt{\xi_1^2+\xi_3^2} \equiv l_{\gamma}$
the linear polarization, and $\xi_0 \equiv 1$.  Since photons have wide
spectra and various polarizations, in general one has to measure
16 two-dimensional luminosity distributions $d{\,^2}L_{ij}/d\omega_1
d\omega_2$,  $dL_{ij} = dL_{\GG} \langle \xi_i \tilde{\xi_j} \rangle$,
where the tilde sign marks the second colliding beam.

Among the 16 cross sections $\sigma_{ij}$, there are three that are most important
as they do not vanish after averaging over the spin states of the final
particles and azimuthal angles, they are~\cite{GKST84,Pak} \vspace{-0.1cm}
\begin{eqnarray}
\sigma^{np} &\equiv& \sigma_{00} =
(\sigma_{\parallel}+\sigma_{\perp})/2=
(\sigma_0+\sigma_2)/2, \nonumber \\
\tau^c &\equiv& \sigma_{22} =
  (\sigma_0-\sigma_2)/2, \nonumber \\
\tau^l &\equiv&   (\sigma_{33} - \sigma_{11})/2 =
(\sigma_{\parallel}-\sigma_{\perp})/2.
\label{cross-sec}
\end{eqnarray}
%\vspace{-0.0cm}
Here the $\sigma_{\parallel}$ and $\sigma_{\perp}$ are the cross sections for
collisions of linearly polarized photons with parallel and orthogonal
relative polarizations and  $\sigma_0$ and $\sigma_2$ are the cross sections
for collisions of photons with  $J_z$ of two photons equal 0 and
2, respectively.

If only these three cross sections are of interest then formula (\ref{dn1}) can
be written as
\begin{equation}
\vspace{-0.1cm}
d\dot{N}_{\GG\ \to X} = dL_{\GG}\; (\sigma^{np} + \langle \xi_2 \tilde{\xi_2}
\rangle \tau^c + \langle \xi_3 \tilde{\xi_3} -
\xi_1 \tilde{\xi_1} \rangle \tau^l)\,.
\end{equation}
Substituting \quad
$\xi_2 \equiv \lambda_{\gamma}$, \quad $ \tilde{\xi_2} \equiv
\tilde{\lambda}_{\gamma}$, \quad $\xi_1 \equiv l_{\gamma} \sin 2\gamma$, \quad
$\tilde{\xi_1} \equiv -\tilde{l}_{\gamma} \sin 2\tilde{\gamma}$,
$\xi_3 \equiv l_{\gamma} \cos 2\gamma$, $\tilde{\xi_3} \equiv
\tilde{l}_{\gamma} \cos 2 \tilde{\gamma}$, and
$\Delta\phi=\gamma -\tilde{\gamma}$ \\(azimuthal angles for linear
polarizations are defined relative
to one $x$ axis), we get
\begin{eqnarray}
d\dot{N} & = & dL_{\GG}( \sigma^{np} + \lambda_{\gamma} \tilde{\lambda}_{\gamma}\;
\tau^c +   l_{\gamma} \tilde{l}_{\gamma} \cos{2\Delta\phi} \;\tau^l)
 \nonumber \\
&\equiv  & dL_{\GG}\;  \sigma^{np} + (dL_0 - dL_2) \tau^c + (dL_{\parallel}
 -dL_{\perp})\,\tau^l   \nonumber \\
& \equiv  & dL_0  \sigma_0 + dL_2  \sigma_2 + (dL_{\parallel} -dL_{\perp})\,
\tau^l   \nonumber \\
&\equiv  & dL_{\parallel}\,  \sigma_{\parallel} + dL_{\perp}\,  \sigma_{\perp}
 + (dL_0 - dL_2)\, \tau^c\,,
\label{dn}
\end{eqnarray}
where
{\small $dL_0 = dL_{\gamma}(1+\lambda_{\gamma} \tilde{\lambda}_{\gamma})/2,\;
dL_2 = dL_{\gamma}(1-\lambda_{\gamma} \tilde{\lambda}_{\gamma})/2,$ \\
$dL_{\parallel} = dL_{\gamma} (1+ l_{\gamma}
\tilde{l}_{\gamma}\cos{2\Delta\phi})/2,\;
dL_{\perp} = dL_{\gamma}(1- l_{\gamma} \tilde{l}_{\gamma}\cos{2\Delta\phi})/2\,.$}

So, one should measure  $dL_{\GG}$, $\langle
\lambda_{\gamma} \tilde{\lambda}_{\gamma}\rangle$, $\langle l_{\gamma}
\tilde{l}_{\gamma}\rangle $ or, alternatively, $dL_0,\, dL_2,\, dL_{\parallel},\,
dL_{\perp}$. If both photon beams have no linear polarization or no
circular polarization, the luminosity
can be decomposed into two parts: $L_0$ and $L_2$, or
$L_{\parallel}$ and $L_{\perp}$, respectively.

For example, for scalar/pseudoscalar resonances ($J=0$)
$\sigma_2=0$, while
$\sigma_{\parallel} = \sigma_0$, \,$\sigma_{\perp} = 0$ for $CP=1$ (scalar)  and
$\sigma_{\perp} = \sigma_0$, \, $\sigma_{\parallel} = 0$ for $CP=-1$ (pseudoscalar).
Then
\begin{equation}
d\dot{N} = dL_{\GG} \,\sigma^{np} (1+\lambda_{\gamma}
\tilde{\lambda}_{\gamma} \pm l_{\gamma} \tilde{l}_{\gamma}\cos{2\Delta\phi})\,.
\end{equation}
In the present work, we investigate the feasibility of studying two-photon production of $C$-even resonance states (charmoniums, bottomoniums, and various exotics in the energy range from 3 to 12 GeV. The cross section for production of narrow resonances in monochromatic non-polarized \GG collisions ($\hslash \equiv c \equiv 1$)
\be
\sigma_{\gamma\gamma \to R}(W) = 8\pi^2(2J+1) \frac{\Gamma_{\gamma\gamma}}{M} \delta(W^2-M^2).
\ee
For broad luminosity spectra and polarized beams, the resonance production rate
\be
    \dot{N} = \frac{d\LGG}{dW_{\GG}}\frac{4\pi^2 (2J+1)\Gamma_{\GG}}{M^2}  \nonumber
 \ee
\vspace{-5mm}
 \be \;\;\;    \times \left( 1 + \frac{\tau^c}{\sigma^{np}} \lambda_{\gamma} \tilde{\lambda}_{\gamma}\;  +  CP \times\frac{\tau^l}{\sigma^{np}} l_{\gamma} \tilde{l}_{\gamma} \cos{2\Delta\phi} \;\right),
\label{ndot}
\ee
  where  $\sigma^{np}, \tau^c, \tau^l$ are defined in (\ref{cross-sec}).

     At the photon collider under discussion,  the degree of polarization in the high-energy part of the spectrum can be close to $100\%$ for the circular and about $85\%$ for linear polarization, which is controlled by the laser polarization.

     For $\lambda_{\gamma} \tilde{\lambda}_{\gamma}=1$, the number of scalars doubles (they are produced only in collisions of photons with the total helicity of zero, with the cross section $\sigma_0$). In the case of $\lambda_{\gamma} \tilde{\lambda}_{\gamma}= - 1$, the total helicity is 2, scalar resonances are not produced, but the number of resonances with $J=2$ almost doubles because it is known that they are produced mostly in the state with  helicity 2  ($\sigma_2 \gg \sigma_0$).
     In the case of linearly polarized $\gamma$-beams,  production of scalars doubles when the linear polarizations of the beams are parallel,  while pseudoscalars, on the contrary, prefer perpendicular linear polarizations.

A nice feature of both \EPEM and \GG\ collisions is the single resonance production of hadrons. At \EPEM\ colliders, resonances with the photon quantum numbers, $J^{PC}=1^{--}$, can be single-produced, which includes the \jpsi and $\Upsilon$ families. On the other hand, two real photons can single-produce $C$-even resonances with the following quantum numbers~\cite{Landau}: $J^P=0^+$, $0^-$, $2^+$, $2^-$, $3^+$, $4^+$, $4^-$, $5^+$, etc., the forbidden numbers being $J^P=1^{\pm}$ and (odd $J)^-$. Therefore, the \GG\ collider presents a much richer opportunity for the study of hadronic resonances.

Resonance production cross sections in \GG\ collisions depend on the total helicity of the two photons, $J_z=0$ or 2. Assuming that the $C$ and $P$ parities are conserved, resonances are produced only in certain helicity states~\cite{Landau}: $J_z=0$ for $J^P=0^\pm$, (even $J)^-$; $J_z=2$ for (odd $J \neq 1)^+$; $J_z=0$ or 2 for $J^P$ = (even $J)^+$. In the experiment, the value of $J_z$ is chosen by varying the laser photon helicities.
\section{Expected $C$-even resonances}
In photon-photon collisions, $C$-even resonances are produced with a wide range of spin and parity values.
The first observation of the $C$-even
resonance, $\eta^{\prime}$ meson at $e^+e^-$ collider was
done by Mark I collaboration in 1979~\cite{EtaP79}. By now, many pseudoscalar~($^1S_0$), scalar~($^3P_0$) and tensor~($^3P_2$) resonances in a wide range of masses
 have been discovered at  $e^+e^-$ colliders
%with initial state radiation (ISR)
in the two-photon fusion process $ e^+e^- \to$ $e^+e^-\gamma^\star \gamma^\star \to e^+e^-X$ by  \babar \; and Belle~\cite{Bphys}, \; CLEO~\cite{CLEO} and BESIII~\cite{BESIII} collaborations.
This process  is dominated by events where both photons are nearly real and both $e^+$ and $e^-$  have very small scattering angles and are not detectable, therefore resonance $X$ and its decay products have small total transverse momentum, which can be used as an experimental sign of the process.
%The allowed J
%P C quantum numbers
%for cc? systems produced via this process are restricted to 0�+ and 2�+.
The cross section of narrow resonance production is proportional to the
  two-photon partial width  $\Gamma_{\gamma\gamma}$ of the resonance thus allowing the measurement
of this quantity at photon colliders, which is the main experimental goal.

\subsection{Heavy quarkonium pseudoscalar, scalar  and tensor states }
\begin{figure}[htb!]
\vspace{-4mm}
\hspace*{-3mm}\includegraphics[width=0.55\textwidth]{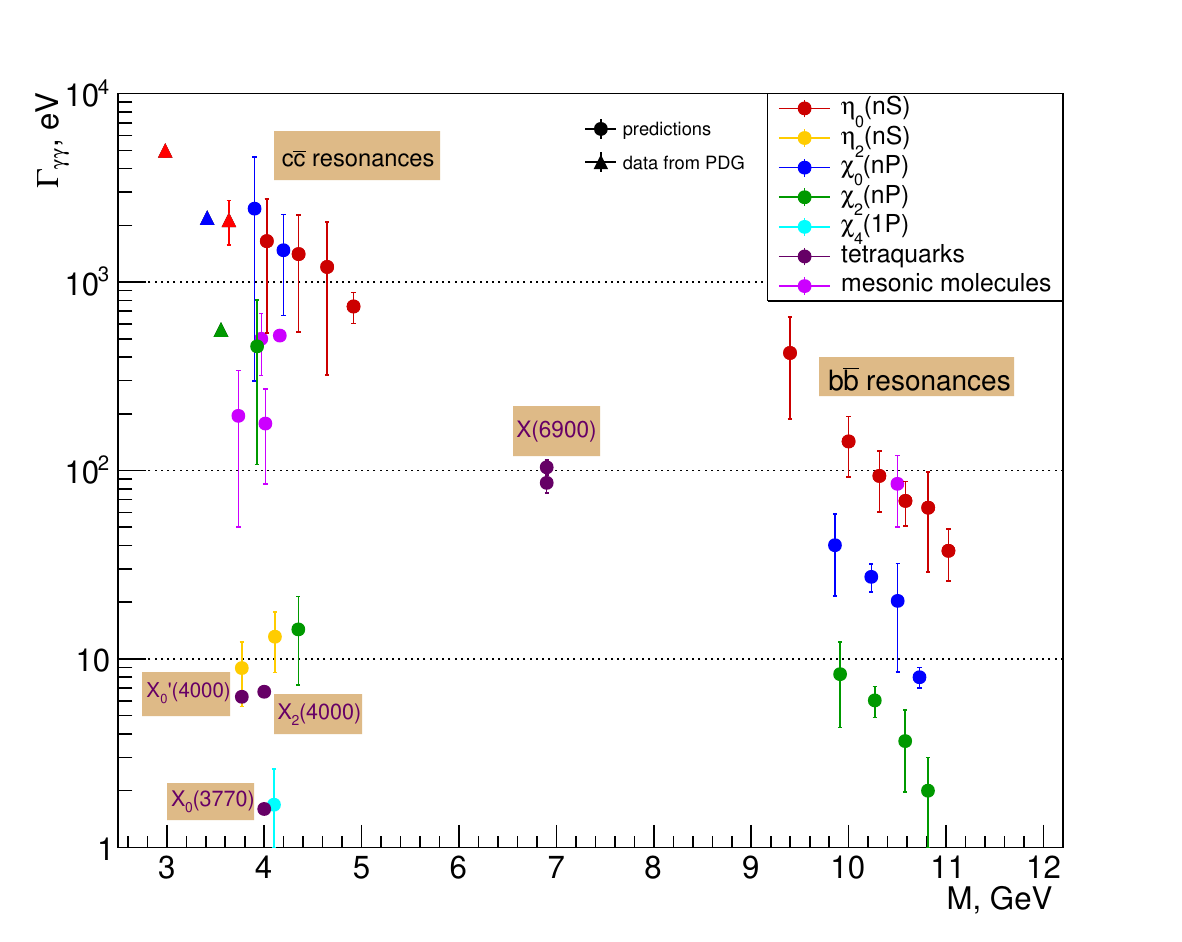}
\caption{Values of the masses and two-photon widths
for various charmonium and bottomonium states from PDG (circles) and various theoretical predictions (triangles), tetraquarks and molecular states
  \label{gammavsm}
}
\end{figure}

In Table~\ref{ggres} and Fig.~\ref{gammavsm}, we list known $c\bar{c}$ and $b\bar{b}$ $C$-even resonances with experimental data from PDG~\cite{pdg} and a summary of theoretical predictions on their
masses and two-photon widths~\cite{chaturC, munz, kher, chatur, li}.

%In Table 1 we compare our predictions for two-photon decay rates of heavy
%quarkonia calculated with the account of both relativistic and radiative corrections
%with previous theoretical calculations and experimental averages from PDG1

%\begin{table}[hbt!]
\begin{table*}[htb]
\caption{
%World-average
Masses and two-photon widths
for various charmonium and bottomonium states from experiment (PDG) and various theoretical predictions~\cite{chaturC, munz, kher, chatur, li} }
\centering
\begin{tabular}{l|l|l|l|l}
\hline
Particle &   Mass (exp.), &  $\Gamma_{\gamma\gamma}$ (exp.), & Mass (pred.), & $\Gamma_{\gamma\gamma}$ (pred.), \\
         & \mevcc        &   keV & \mevcc & keV
\\[1mm]\hline
\multicolumn{5}{l}{$c \bar{c}$ resonances} \\
\hline
$\eta_{c0}(1S)$ & 2983.9 $\pm$ 0.5 & 5.0 $\pm$ 0.4 & 2976 -- 3014 & 1.12 - 9.7 \\
$\eta_{c0}(2S)$ & 3637.5 $\pm$ 1.1 & 2.14 $\pm$ 0.57 & 3584 -- 3707 & 0.94 - 5.79 \\
$\eta_{c0}(3S)$ & -- & -- & 3991 -- 4130 & 0.30 -- 4.53 \\	
$\eta_{c0}(4S)$ & -- & -- & 4425 -- 4384  & 0.50 -- 2.44 \\
$\eta_{c0}(5S)$ & -- & -- & 3991 -- 4130 & 0.42 -- 2.21 \\	
$\eta_{c0}(6S)$ & -- & -- & 4425 -- 4384  & 2.16 -- 3.38 \\
$\eta_{c2}(1S)$ & -- & -- & 4425 -- 4384  & 0.009 -- 0.013 \\
$\eta_{c2}(2S)$ & -- & -- & 4425 -- 4384  & 0.0072 -- 0.0202 \\
$\eta_{c4}(1S)$ & -- & -- & 4425 -- 4384  & (0.3 -- 3) $\cdot 10^{-4}$ \\
$\chi_{c0}(1P)$ & 3414.71 $\pm$ 0.30 & 2.20 $\pm$ 0.16 & 3404 -- 3474 & 1.18 -- 2.62 \\
$\chi_{c0}(2P)$ & 3921.7 $\pm$ 1.8 & -- & 3901 $\pm$ 1 & 0.64 -- 2.67 \\	
$\chi_{c0}(3P)$ & -- & -- & 4197 $\pm$ 3 & 0.74 -- 2.77 \\	
$\chi_{c0}(4P)$ & 4704$^{+17}_{-20}$ & -- & 4700 $\pm$ 2 & 1.24 -- 1.24 \\	
$\chi_{c2}(1P)$ & 3556.17 $\pm$ 0.07 & 0.56 $\pm$  0.03 & 3488 -- 3557 & 0.22 -- 1.72 \\
$\chi_{c2}(2P)$ & 3913.17 $\pm$ 0.0.07 & -- & 3927 $\pm$ 26 & 0.27 -- 0.58 \\
$\chi_{c2}(3P)$ & 4350 $\pm$ 7 & -- & 4280 -- 4427 & 0.014 -- 1.49 \\
$\chi_{c2}(4P)$ & -- & -- & 4614 -- 4802 & 1.69  \\
$\chi_{c3}(1P)$ & -- & -- & 4000 & 0.00044 -- 0.003 \\
$\chi_{c4}(1P)$ & -- & -- & 3990 & 0.00031 -- 0.0012 \\
\hline
\multicolumn{5}{l}{$b \bar{b}$ resonances} \\
\hline
$\eta_{b0}(1S)$ & 9398.7$\pm$2.0	& -- &	9391  & 0.46 -- 0.86 \\
$\eta_{b0}(2S)$ & -- & -- & 9999  & 0.07 -- 0.26 \\
$\eta_{b0}(3S)$ & -- & -- & 10315  & 0.04 -- 0.09 \\
$\eta_{b0}(4S)$ & -- & -- & 10583  & 0.05 -- 0.76 \\
$\eta_{b0}(5S)$ & -- & -- & 10816  & 0.04 -- 0.12 \\
$\eta_{b0}(6S)$ & -- & -- & 11024  & 0.03 -- 0.05 \\
$\eta_{b2}(1S)$ & -- & -- & 10130  & (2.83 -- 5.13)$\cdot 10^{-5}$  \\
$\eta_{b2}(2S)$ & -- & -- & 10430  & (5.23 -- 96.2)  $\cdot 10^{-5}$ \\
$\eta_{b4}(1S)$ & -- & -- & 10510  & (1.6 -- 7.2) $\cdot 10^{-8}$ \\
$\chi_{b0}(1P)$ & 9859.44 $\pm$ 0.52 & -- &	9849 &	0.021 -- 0.069  \\
$\chi_{b0}(2P)$ &  10232.5 $\pm$0.6 & -- &	10226 &  0.022 -- 0.027 \\	
$\chi_{b0}(2P)$ & -- & -- & 10503  & 0.012 -- 0.037 \\
$\chi_{b0}(4P)$ & -- & -- & 10727  & 0.08 \\
$\chi_{b2}(1P)$ & 9912.21 $\pm$ 0.40 &	--	&	9900 & 0.005 -- 0.016 \\
$\chi_{b2}(2P)$ & 	10268.65$\pm$0.54	& -- &	10257 & 0.004 -- 0.006 \\	
$\chi_{b2}(3P)$ & 	10524.0$\pm$0.8	& -- &	10578 & 0.002 -- 0.006 \\
$\chi_{b2}(4P)$ & 	--	& --	&	10814 & 0.002 \\
$\chi_{b4}(1P)$ &	--	& --	&	10350 -10390 & (0.58 -- 1.94) $\cdot 10^{-6}$ \\
\hline
\end{tabular}
\label{ggres}
\end{table*}

%Two-photon widths can be calculated in  perturbative QCD.
Two-photon widths are successfully predicted with nonrelativistic quark models~\cite{chatur}.
In the nonrelativistic limit, two-photon widths of mesons
are proportional to the squa\-re of the wave function or its derivative at the origin.
However, relativistic effects are important, especially for charmonium, and  modify this relation~\cite{chaturC, munz, kher,li}.
The first-order correction is proportional to the QCD coupling $\alpha_s$, which is estimated to be
$\alpha_s(m_b)$ = 0.18 for bottomonia and $\alpha_s(m_c)$ = 0.26
for charmonia, respectively~\cite{ebert}.
%So  the two-photon
%decays of pseudoscalar, scalar and tensor states of charmonium and
%bottomonium are calculated in the framework of the relativistic quark model~\cite{ebert}.

Another way to study non-perturbative QCD is an lattice QCD~\cite{okamoto, chen, Liao, ychen, choe}, which is quantum field theory defined on  discrete Euclidean space-time.

Two-photon decay widths for scalar
and pseudoscalar charmonia were recently estimated to be about 1~keV~\cite{ychen} which is smaller than the experimental values.

Besides quark-antiquark pairs mesons the quark model assumes the existance of exotic multiquark hadrons with more complex
internal structures.
%for example such as glueballs, tetraquark or molecular states.
Neutral mesons with exotic properties, namely, $X$- and $Y$-states in the mass range from 3.8 to 7.0~\gevcc, have been discovered experimentally.
Different interpretations have been proposed for those
%experimentally observed and predicted states from table~\ref{ggres} and also for new multiquark states  that are briefly discussed below.
resonances summarized in~\cite{universe}, such as  tetraquarks, molecular states, quark-gluon hybrids, hadro-quarkonia, kinematic threshold effects, or mix states.
Feasibility of observaion multiquark states  in $\gamma\gamma$ collisions is discussed below.
\subsection{Tetraquarks}
The simplest multiquark system is a tetraquark, which consists of two quarks and two
antiquarks, is color- and charge-neutral, and has spin not equal to 1.
One possible way to check for the  existence of tetraquarks is to find a complete flavor-spin multiplet
such as the standard quarkonium families.
Scalar and tensor states are expected to be produced in two-photon collisions,
although their two-photon widths are expected to be less than 1~keV~\cite{giacosa}.
A lot of tetraquarks that can be produced in $\gamma \gamma$ collisions with masses from 3 to 12 \gevcc
are  predicted in the relativistic quark model
based on the quasipotential approach in the recent work~\cite{universe}.
In those calculations tetraquarks were assumed to have two or four heavy quarks
 and a diquark-antidiquark picture of heavy tetraquarks was used.

A narrow resonance around 6.9~\gevcc in the invariant mass spectrum of $J/\psi$ pairs  was found by LHCb collaboration~\cite{x6900} and was called $X$(6900).
 Its mass and width were measured to be $M_X\! =\! 6886\! \pm\! 2$~\mevcc and $\Gamma_X\! =\! 168\! \pm\! 102$~MeV,
while its quantum numbers can be $0^{++}$ or $2^{++}$.
This resonance can be interpreted as a $cc\bar{c}\bar{c}$ compact state.
Using the vector\; meson\; dominance\; model\; in\; the\; assumption\; of\; its\; strong coupling to a di-$J/\psi$ final
state the X(6900) two-photon width has been estimated as 104~eV for $J^{PC}=0^{++}$ and
86~eV for $J^{PC}=2^{++}$~\cite{tetra1}.
%formulas

Scalar and tensor tetraquarks $cc\bar{q}\bar{q}$ exist in the diquarkonium model
but have  not been observed yet in any experiment.
Two states  with quantum numbers  $J^{PC}=0^{++}$ and one with  $J^{PC}=2^{++}$
are predicted by the diquark-antidiquark model with the dominated $cq$ interaction,
and their masses are 3770~\mevcc, 4000~~\mevcc and 4000~~\mevcc, called $X_0(3770)$,\, $X_0^{\prime}(4000)$ and $X_2(4000)$, respectively~\cite{maiani}.
The partial $\gamma\gamma$ widths of those tetraquarks are predicted to be 6.3~eV, 6.7~eV and 1.6~eV, respectively~\cite{tetra1}. Experimental search for these states is an important test of the
diquark--antidiquark picture of heavy tetraquarks.
\vspace{-3mm}
\subsection{Mesonic molecules}
\vspace{-3mm}
Hadronic molecules are bound states of two or more mesons.
Particles with masses close to the sum of masses of two other mesons, on one hand, and away from the predictions of the quark model
on the other
%are often treated as mesonic molecules.
are often considered to have  a possible molecular structure.
The most famous experimental candidate for a mesonic molecule is $X$(3872) resonance,
which is considered to be  $D^0\bar{D}^{\star 0}$ molecule~\cite{ufn}.
Other  heavy meson  candidates with mass greater than 3~\gevcc considered to have a molecular structure
are  $X$(3915)~\cite{babar, belle}, $Y$(3940), $Y$(4140) and $Y$(4660)~\cite{y4660}.
Identification of an observed resonance as a mesonic molecule is based
not only on its mass and quantum numbers but also on the process in which the resonance was found.
For predictions, the theory of
electromagnetic interactions is usually used. So, properties of resonance produced in two-photon collisions
provide information about its nature.

Partial two-photon widths calculated in the framework of a phenomenological Lagrangian approach  of $D\bar{D}$, $D_s\bar{D}_s$, $B\bar{B}$ molecules
are expected in the range 0.1--2.8~keV~\cite{mol1}.
%for pseudoscalar or scalar charmed and bottom
%mesons~\cite{mol1}. In particular,  r
Radiative \; widths \;
of\; the\; molecules\; $Y$(3940) = ${D^{\star}\bar{D}^{\star}}$ \; and  $Y$(4140) =${D^{\star}_s \bar{D}^{\star}_s}$ are about 1~keV~\cite{mol2}.

\vspace{-2mm}
\subsection{Glueballs}
\vspace{-2mm}
Glueballs predicted by QCD are color-neutral states that consist only of gluons.
Gluons inside a glueball can self-interact but gluons remain stable, except the heaviest states that decay into lighter glueballs.
%quarkless massive bound states
%Many particles of this type are suggested in theory.
Theory suggests a rich spectrum of glueballs.
Their existence is compatible with recent experimental data,
and several exotic meson candidates have been interpreted as glueballs:
$f_0(1370)$, $f_0(1500)$, $f_0(1710)$, $f_J(2220)$ and others.
%, most famous one searched by CLEO~[].
The challenge is
%not to find a glueball
to identify observed particles as glueballs. The situation is complicated by the
lack of know\-ledge on the glueball nature
and possible
mixing of glueballs with standard quark model states.

Glueball production in two-photon collisions is a unique process that can
clearly  distinguish a tensor glueball from a tensor meson~\cite{glue1}.
Gluons do not participate in electromagnetic interactions. Two-photon widths of  glueball states are significantly smaller in comparison with two photon widths of ordinary quarkonia~\cite{glue2}.
The advantage is that two-photon width is model-independent
in contrast with other glueball properties.
The expected two-photon width is 1--10~eV.

Glueballs are predicted in lattice QCD calculations.
The mass of the first excited glueball in the tensor channel is estimated using lattices
to be  $3320 \pm 20 \pm 160$~\mevcc~\cite{glue3}.
States with quantum numbers and masses
$$
J^{PC}=2^{-+},\; m_G=3040 \pm 40 \pm 150~\mevcc,
$$
\vspace{-6mm}
$$
\hspace*{-1mm}J^{PC}=3^{++}, \;m_G=3670 \pm 50 \pm 180~\mevcc
$$
are predicted with the improved technique~\cite{glue4}.

\section{\GG\ collider}

The parameters of the \GG\ collider based on the 17.5 GeV electron linac of the European XFEL is described in ref.~\cite{telnov-gg12}.
The maximum energy of scattered photons
\be
\omega_m \approx \frac{x}{x+1}E_0, x = \frac{4E_0\omega_0}{m^2c^4}=
 19\left[\frac{E_0}{\tev}\right]
\left[\frac{\mum}{\lambda}\right].
\label{x}
\ee

For $E_0=17.5$ \gev and the laser wavelength $\lambda=0.5$ \mum,  $x=0.65$, $\omega_m/E_0=x/(x+1) \approx 0.394$,  $W_{\GG, \rm max} \approx 13.3$ \gevcc, with a peak at 12 \gevcc, which covers the region of \bbbar resonances. The peak energy can be varied by adjusting the electron beam energy. The thickness of the laser target is taken to be equal to one scattering length for electrons with the initial energy. The required flash energy is about 3 J. We consider both unpolarized (as currently available at the European XFEL) and $80\%$ longitudinally polarized electron beams. The laser beam should be circularly polarized, $P_c= \pm 1$, when circularly polarized high-energy photons are needed. Collisions of linearly polarized photons would also be of interest for physics; for that, linearly polarized laser beams should be used. The degree of circular polarization in the high-energy part of the spectrum can be close to $100\%$ (for any $x$)  and about $85\%$ for linear polarization (for $x=0.65$).

\begin{figure*}[h]
\centering
\includegraphics[width=7.5cm,height=6.5cm,angle=0]{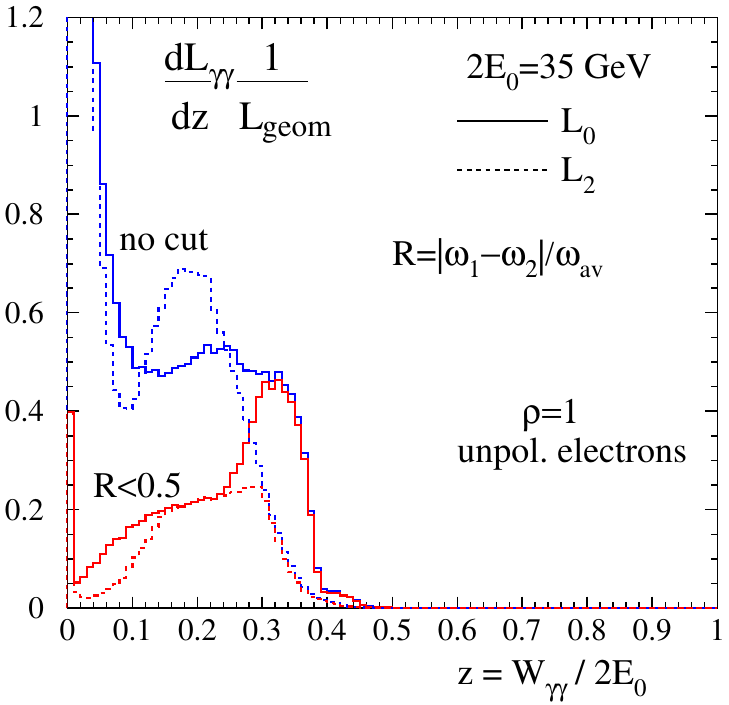} \hspace*{1cm} \includegraphics[width=7.5cm,height=6.5cm,angle=0]{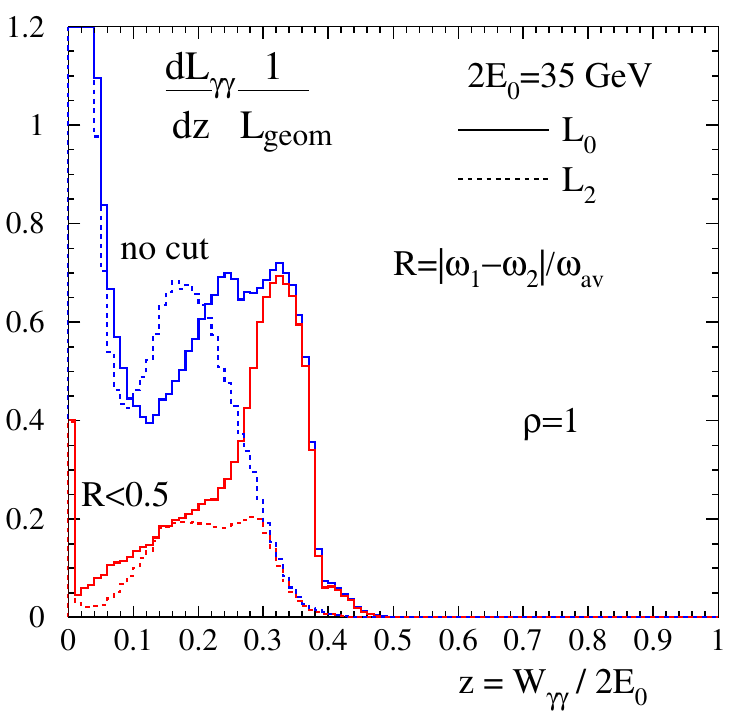}
%\vspace*{-0.2cm}
\caption{\GG\ luminosity distributions vs the invariant mass $W_{\GG}$: \emph{(left)} unpolarized electrons; \emph{(right)} longitudinal electron polarization $2\lambda_e=0.8$ (80$\%$). In both cases the laser photons are circularly polarized, $P_c=-1$. Solid lines are for the total helicity of the two colliding photons $J_z=0$, dotted lines for $J_z=2$. Red curves are luminosities with a cut on the longitudinal momentum.} \label{pol-np}
\end{figure*}

The \GG\ luminosity spectra for non-polarized and longitudinally polarized electrons are shown in Fig.~\ref{pol-np}.
The spectra are decomposed into states with the total helicity  of the colliding photons $J_z=0$ or $2$; the total luminosity is the sum of the two spectra. Also shown are the luminosities with a cut on the relative longitudinal momentum of the produced system that suppresses boosted collisions of photons with very different energies. Luminosity distributions similar to those in Fig.~\ref{pol-np} but for various distances $b$ between the conversion and interaction points are given in ref.~\cite{telnov-gg12}. As the distance increases, the luminosity spectra become more monochromatic at the cost of some reduction in luminosity.

For the study of resonances, when the invariant mass is determined by the detector, the maximum luminosity is needed; therefore a small distance is preferable, as in Fig.~\ref{pol-np}, where $\rho=b/\gamma \sigma_y=1$ and corresponding $b = 1.8$ mm.
\begin{figure}[htbp]
\vspace{-0.5cm}
\centering
\hspace*{-0.2cm}\includegraphics[width=0.47\textwidth,height=6.5cm]{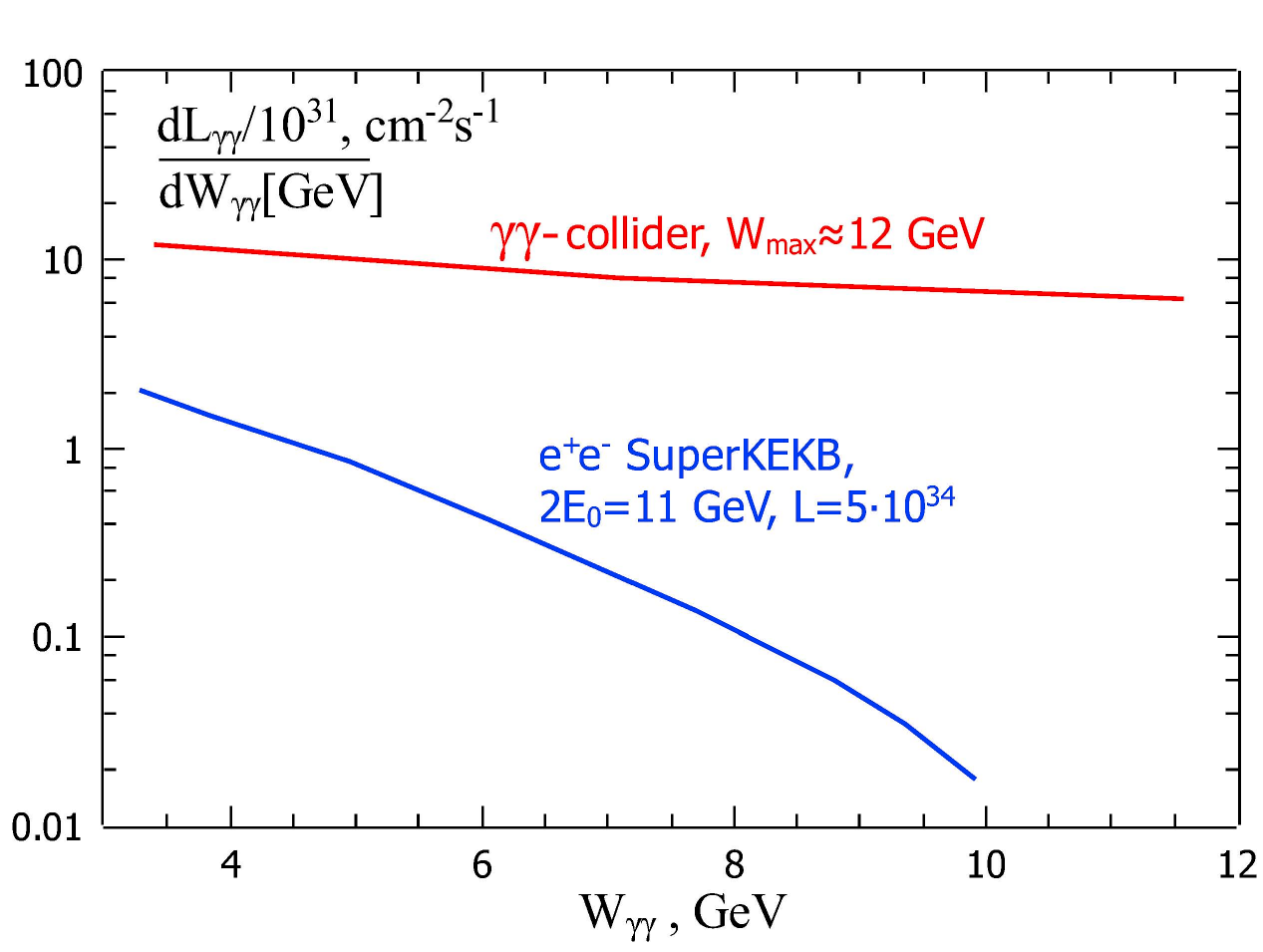}
%\vspace{-0.3cm}
\caption{Comparison of \GG\ luminosities at the photon collider and Super-KEKB.}
\vspace{-0.2cm}
\label{lumkek}
\end{figure}
The geometric electron-electron luminosity at the nominal energy of 17.5 GeV $\LEE = 1.45 \cdot 10^{33}\,\cms$ (determined by the beam emittances and proportional to the energy), $\LGG(z>0.5z_m) \approx 2\cdot 10^{32}\,\cms$ ($\propto \LEE$). The resonance production rate is proportional to $d\LGG/d\WGG$ at the peak of the luminosity distribution. In Fig.~\ref{lumkek} it is compared with that at  SuperKEKB in $\gamma^*\gamma^*$ collisions for $2E_0=11$ GeV and $\LEE = 5\cdot 10^{34}\,\cms$ . At present, $L_{\rm max} \sim 4.5\cdot 10^{34}\,\cms$ at SuperKEKB, the planned value (in the year 2028) is $L_{\rm max} \sim 8\cdot 10^{35}\, \cms$. In any case, the photon collider is beyond competition in the \bbbar energy region.
\section{Suppression of hadronic background}
Below we study the possibility of detecting narrow $C$-even resonances in the process $\GG \to R$ at $W=$4--10 \gevcc. The selection of resonances requires registration of all final-state particles. The effective cross section of resonance production is proportional to $\Gamma_{\GG}/M_R^2$ (\ref{ndot}). For bottomonium (\bbbar) states, this value is two orders of magnitude smaller than for charmonium (\ccbar) states.  At the same time, the cross section of the background  $\GG\ \to hadrons$ process in this region is almost constant, $\sigma_{\GG \to \rm{hadr}}\sim 350$ nb. For example, the first candidate for studying is $\eta_b(9400)$ with $\GGG \sim 0.5$ keV (the largest in this mass region). The number of hadronic events in the  resonance mass region $\pm \sigma_M$ ($\sigma_M \approx 50$ \mevcc) will be about 230 times greater than the number of resonances.
\begin{figure}[htbp]
\centering
\includegraphics[width=0.45\textwidth]{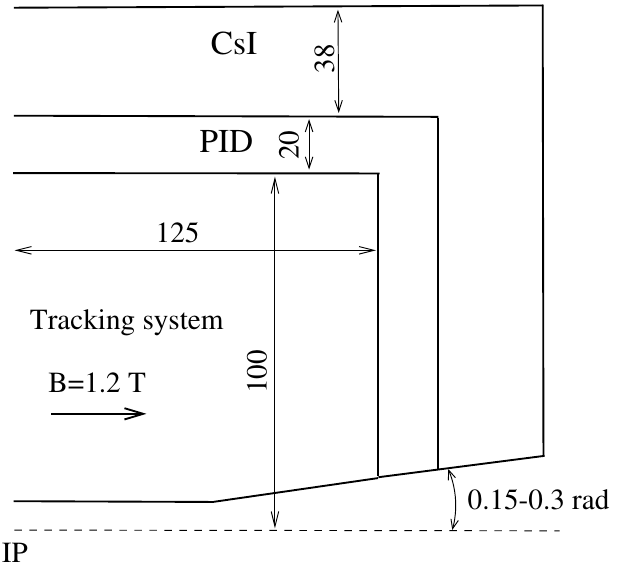}
\caption{The layout of the detector (a quarter of the inner part).}
\label{detector}
\end{figure}
In the present study, we carefully consider this problem, trying to suppress the background and to maximize the significance of the resonances, i.e. to increase the value  $S/\sqrt{B}$.

The procedure is the following. We simulate resonances and hadrons at several invariant masses, from 4 to 10 \gevcc, 100000 events of resonances and hadrons at each point.  Resonances and hadrons are generated by PYTHIA~\cite{pythia}. Re\-so\-nances are modeled as $\eta_b$, but with changed masses. Hadrons were modeled with a mass spread of 10\% (similar to the width of the high-energy peak at the \GG collider) at the same average invariant mass as the resonance under study. It is assumed that the peak in the luminosity distribution over invariant masses coincides with the resonance mass. If a hadronic event has passed all the selection conditions but its reconstructed mass is more than 20 \% lower than the average peak mass, then this is no longer a background for the studied resonance. There are quite a few such cases; the requirement of a small total transverse momentum cuts off all events with lost particles.

 Simulated particles passed the detector (described by GE\-ANT-4) shown in Fig.~\ref{detector} (only the elements used for analysis are shown). It has a tracking system, a particle identification system (PID) and a CsI EM calorimeter. At this stage (it is just first look), we did not make a complete reconstruction of the event, but rather assigned a certain momentum and angular resolution to the particles that passed through the detector.

However, this is not just a geometric simulation: GEANT simulates interactions with a vacuum chamber and particle decays, which affects the reconstruction efficiency. A charged particle  considered registered if it passes at least 55 cm through the track system. Reconstruction of resonances requires the registration of all (detectable) particles. The average particle multiplicity is about 17 at 10 GeV. Particles that are undetectable (e.g., neutrinos) or that could mess up the mass resolution ($n,\bar{n},\bar{p},K_L$) were simply removed from the events (after which the event did not pass the selection criteria or had a displaced mass). Removal of these particles has reduced the reconstruction efficiency by a factor of 1.7 and 3.4 for resonances with masses 4 and 10 \gevcc, respectively.

Thus, good events consist only of $\pi^{\pm}, K^{\pm}, e^{\pm}, \mu^{\pm} $ and photons (events with protons are bad because they always contain $\bar{p}$ or $\bar{n}$). Calculating the invariant mass, we assumed that electrons and positrons are completely identified ($m=m_e$). For muons,  we assigned the pion mass.  Kaons were considered identified ($m=m_K$) at $pc<0.5$ GeV (by $dE/dx$ in the tracking system) or when they crossed the PID system (it may be some kind of an aerogel system), otherwise they were assigned the the pion mass. If charged kaons are not identified at all, then the efficiency (the number of resonances in a narrow peak) will decrease by a factor of 1.3 times (at 10 \gevcc).

Detector parameters are the following: the minimum angle is 0.15 rad (0.30 was checked as well), the magnetic field $B= 1.2$ T, the tracking resolution
$$(\sigma_{p_{\perp}}/p_{\perp})^2=(2\cdot 10^{-3} p_{\perp}[{\rm GeV/c}])^2 + (3\cdot 10^{-3}/\beta)^2,$$
the e.m. calorimeter resolution
$$\sigma_E/E = 0.025/\sqrt{E[\rm GeV]}.$$
The invariant mass resolution is almost completely determined by the energy resolution of the calorimeter. The contribution of any typical photon angular resolution is negligible, since the system is created almost at rest ($P_{\|}c<0.1\,W$).

   Below we choose the selection criteria to optimally suppress the hadron background and find the detection efficiency for resonances and the hadron background. Consideration is made for a minimum detector angle of 0.15 rad. At the end, we present the results for 0.3 rad. The difference in efficiencies turns out to be insignificant since  events with a large sum of transverse particle momenta are selected to separate resonances from the background.

\begin{figure*}[htbp]
\centering
\includegraphics[width=0.5\textwidth]{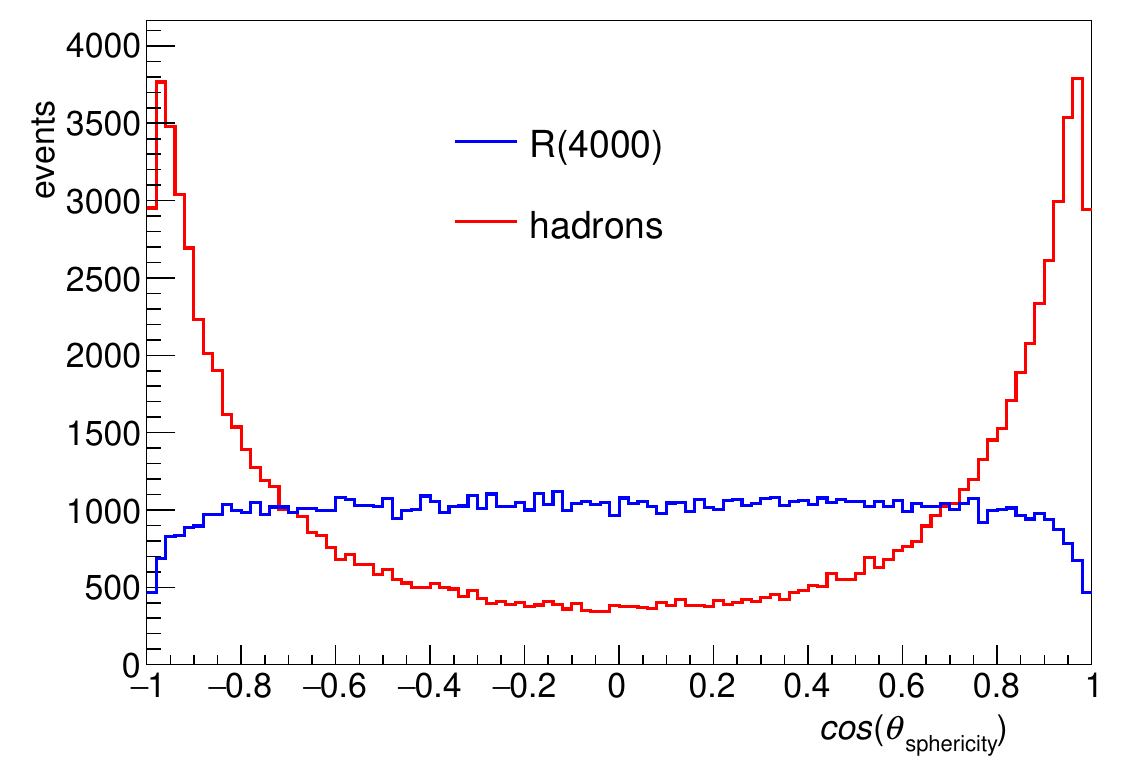}\includegraphics[width=0.5\textwidth]{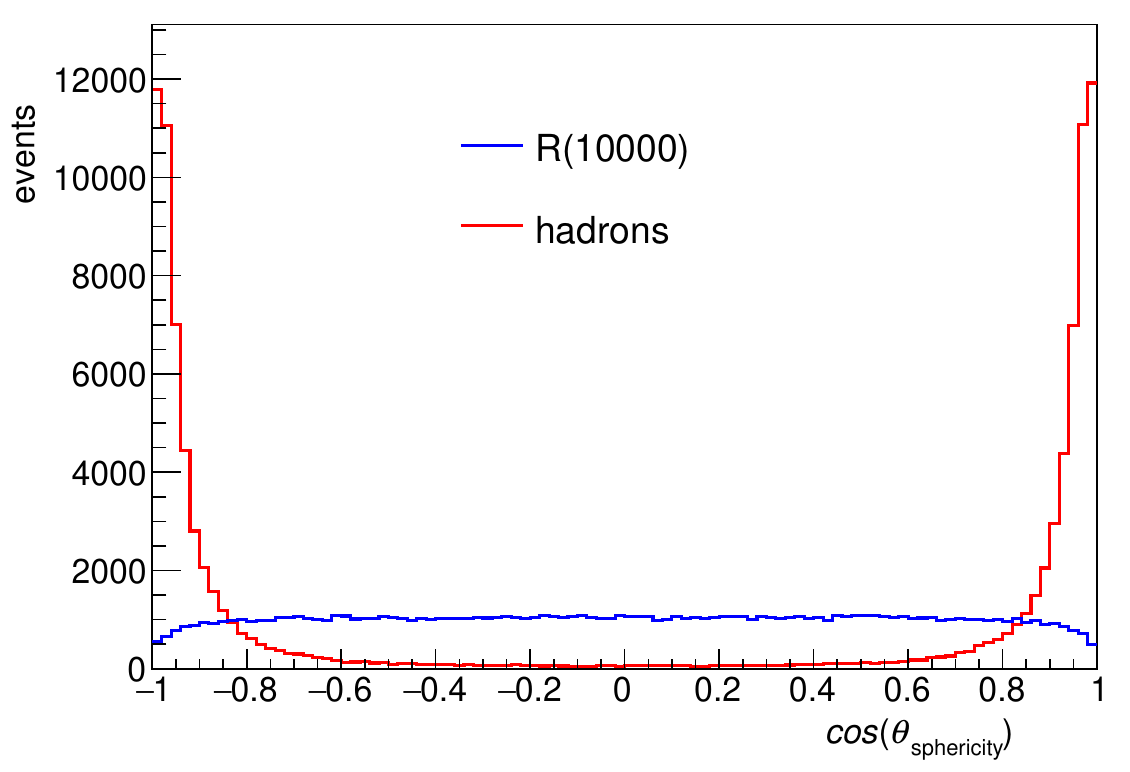}
\caption{The distributions of resonance and hadronic events on the sphericity angle for $W=$ 4 and 10 GeV/$c^2$ }
\label{spher-angle}
\end{figure*}

  The sphericity angle distributions for the resonance and hadronic events are shown in Fig.~\ref{spher-angle}. It can be seen that the hadronic background at $W=10$ \gevcc is pressed to the axis more strongly than at $W=4$ \gevcc; these difference can be used to suppress hadrons. Below we consider the selection criteria based on the differences in the angular distribution of particles from the decay of resonances (with $J=0$) and the hadronic background.

\begin{figure*}[htbp]
\centering
\includegraphics[width=0.5\textwidth]{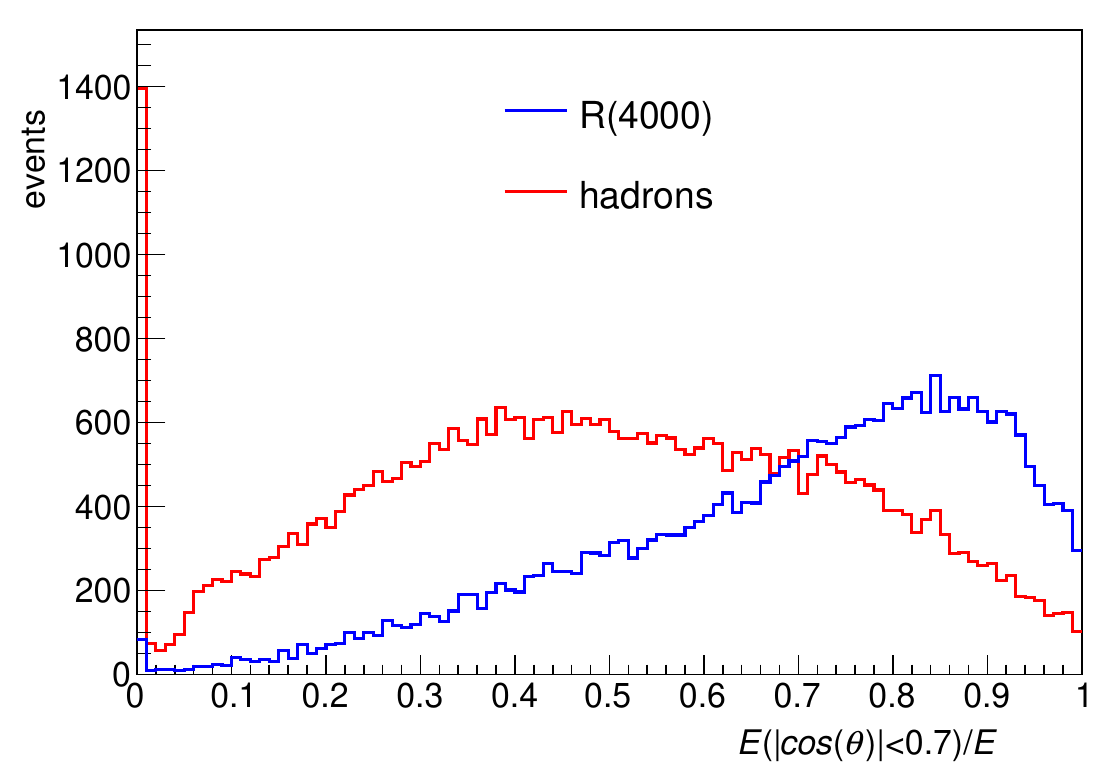}\includegraphics[width=0.5\textwidth]{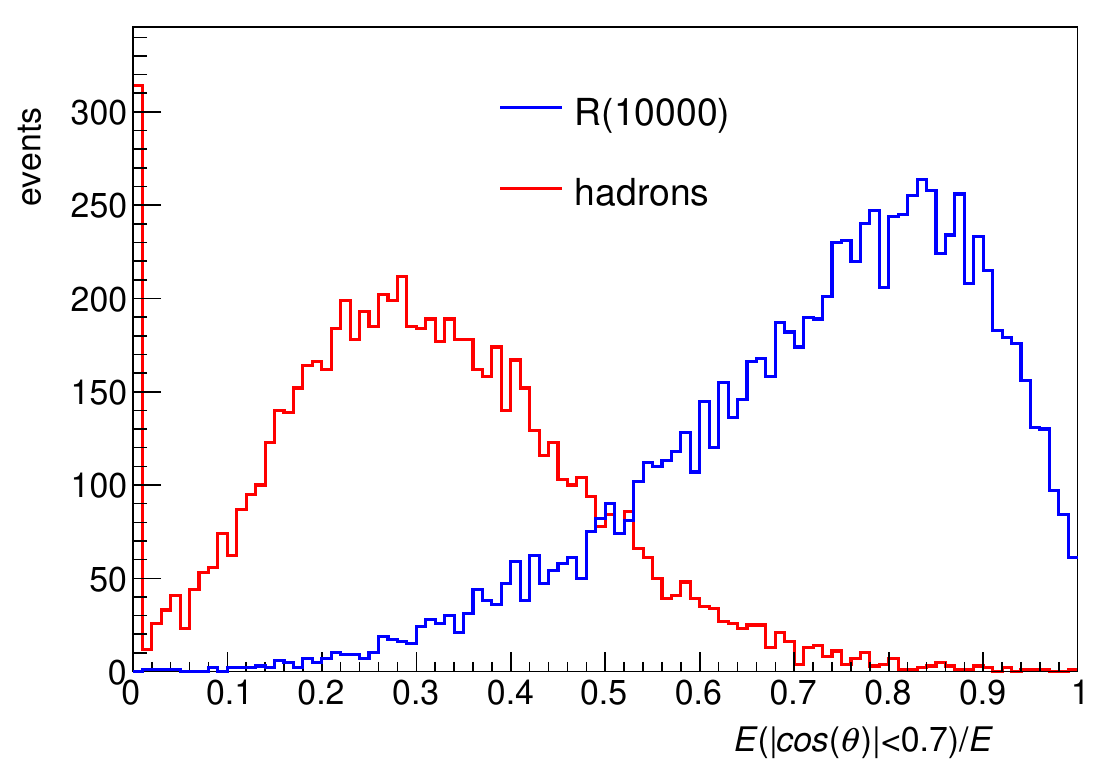}
\caption{The distributions of the parameter  $E(|\cos{\theta}|<0.7)/E$ for resonances with M = 4 and 10 GeV/$c^2$ and corresponding hadronic background. }
\label{cos07}
\end{figure*}

\begin{figure*}[htbp]
\centering
\includegraphics[width=0.5\textwidth]{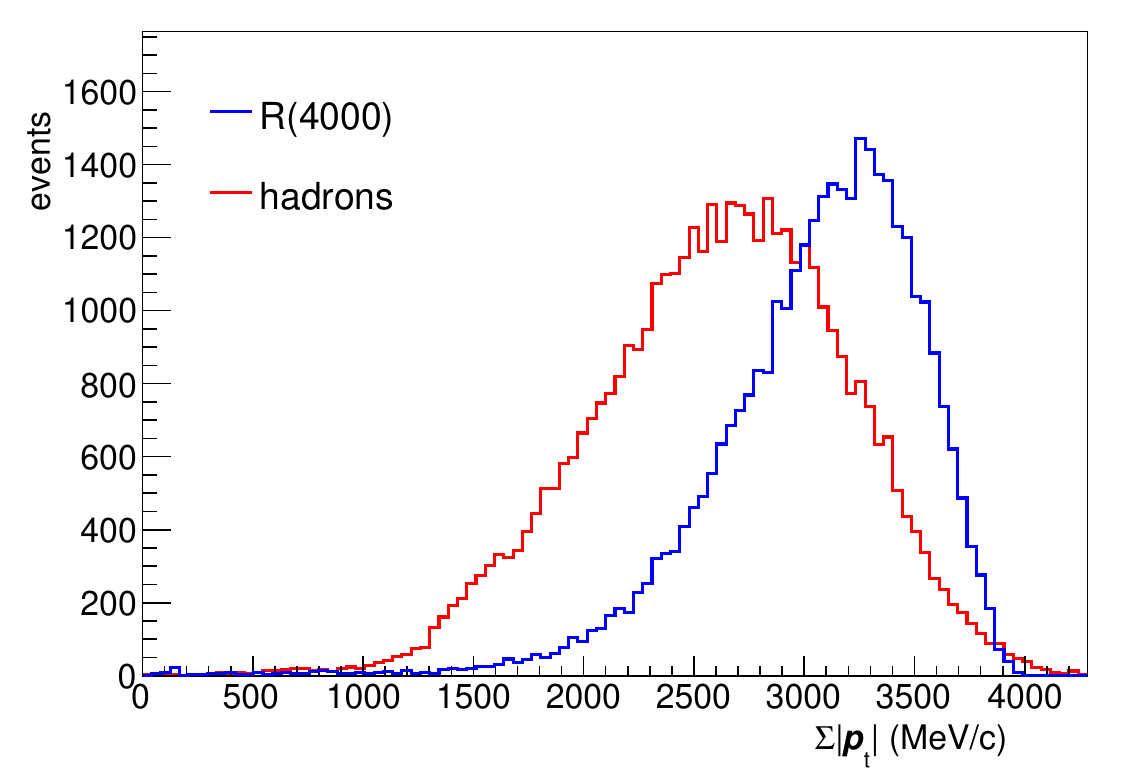}\includegraphics[width=0.5\textwidth]{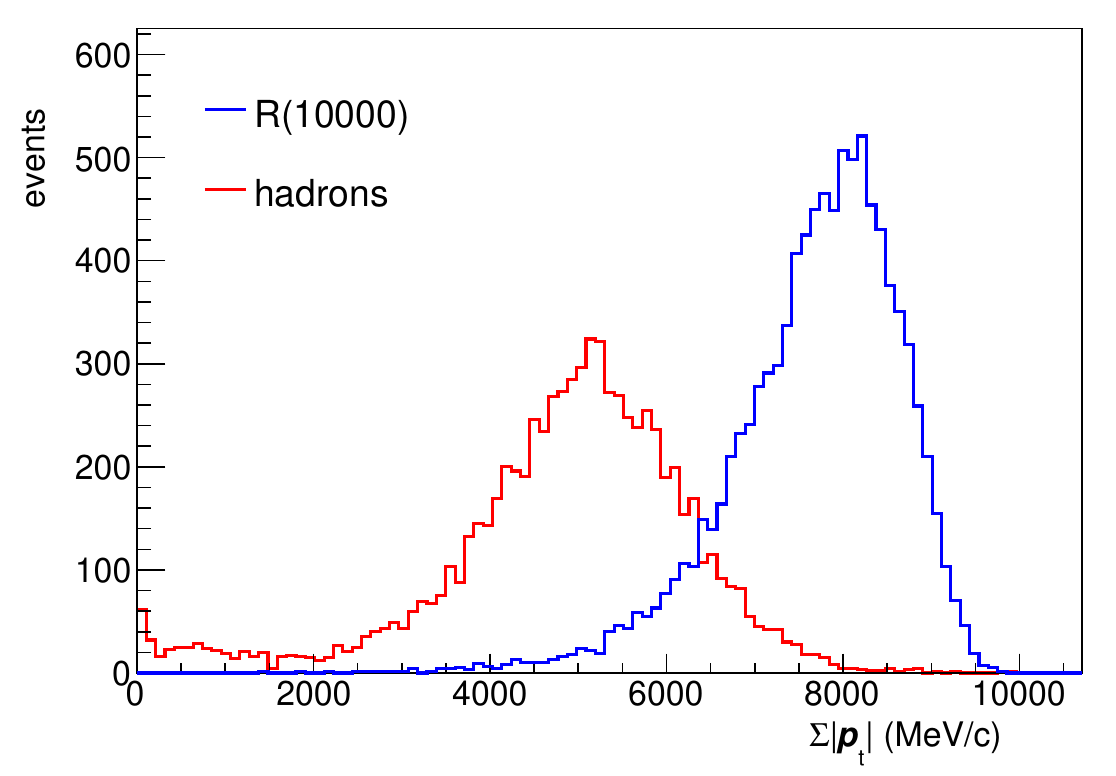}
\caption{The distribution  $\Sigma |p_t|$ for resonances with M = 4 and 10 GeV/$c^2$ and hadronic background.}
\label{sum-mod-pt}
\end{figure*}

    One of the conditions is based on the ratio of the sum of the particles' energies in the detector at an angle larger than some $\theta_{\rm min}$ to the total energy in the detector. For $W=10$ \gevcc the optimum angle is about $|\cos{\theta}|=0.7$. The distributions of the ratio $E(|\cos{\theta}|<0.7)/E$ is shown in Fig.~\ref{cos07}. We have found that the optimal value of this ratio for hadron suppression is about 0.7. This is the first constraint for the separation of resonances
   \be
     1) \;\;\;\;\; E(|\cos{\theta}|<0.7)/E <0.7.
   \label{1)}
    \ee
   The distributions on $\Sigma |p_t|$ are shown is Fig.~\ref{sum-mod-pt}. In terms of separating power, it is comparable to the previous cut. For the selection of a resonance with the mass $M$, we require $\Sigma |p_t| > 0.75 Mc$; this is the second constraint:
   \be
   2) \;\;\;\;\;\Sigma |p_t| > 0.75\,Mc.
   \label{2)}
   \ee
    The constraints 1) and 2) strongly correlate; nevertheless, together they give a slightly better result.

\begin{figure*}[htbp]
\centering
\includegraphics[width=0.5\textwidth]{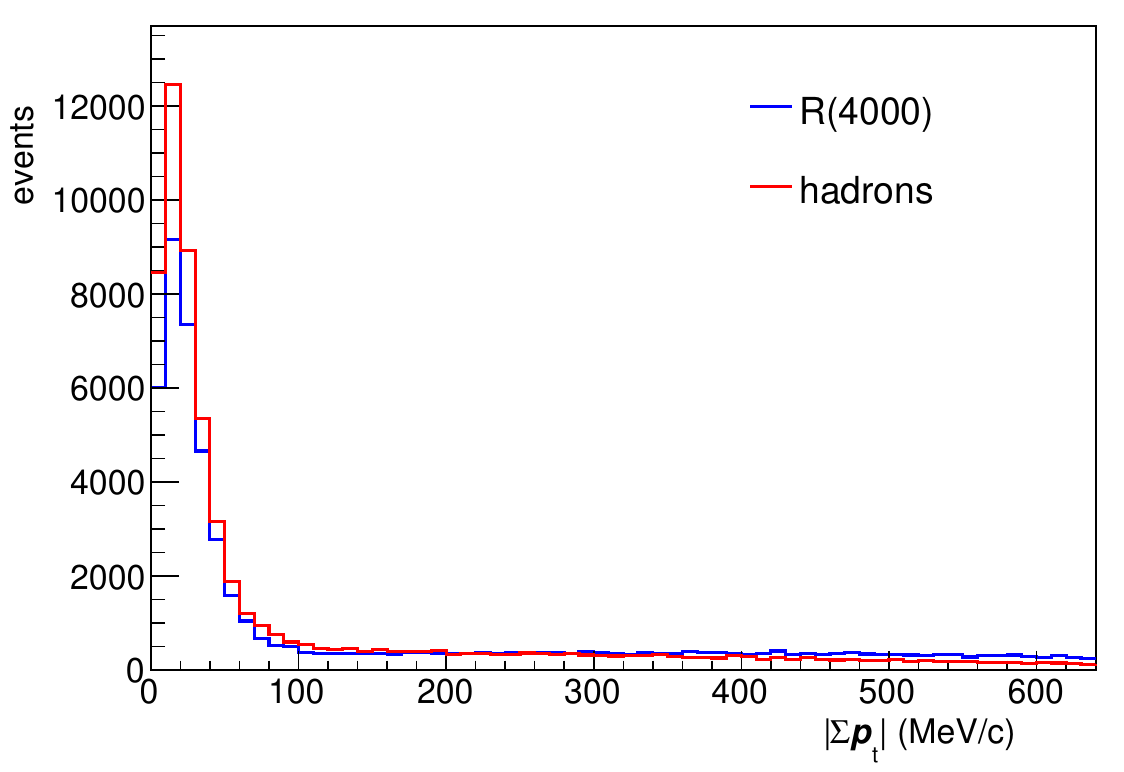}\includegraphics[width=0.5\textwidth]{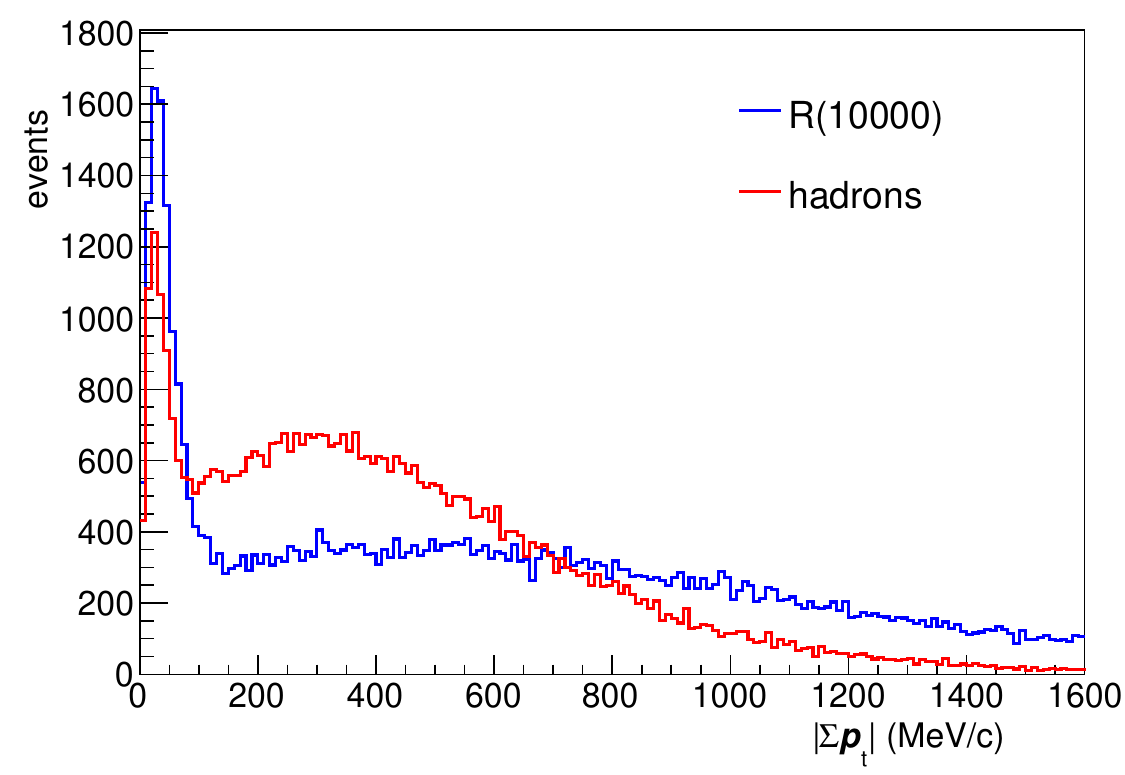}
\caption{Distributions on $|\Sigma \vec{p}_t|$ in the detector (without any other cuts)  for resonances with $M$ = 4 and 10 GeV/$c^2$ and hadronic background.}
\label{sumpt}
\end{figure*}

    The distribution for all events (without any cuts) of the total\; transverse\; momentum\; $|\Sigma \vec{p}_t|$ \;of detected\; particles\; is shown in Fig.~\ref{sumpt}. Only events with a small $|\Sigma \vec{p}_t|$ (this indicates that all particles were registered by the detector) are suitable for observing narrow resonances. This defined our third cut:
    \be
    3) \;\;\;\;\;|\Sigma \vec{p}_t|<100\,\, {\rm MeV}/c.
    \ee

\begin{figure*}[htbp]
\centering
\includegraphics[width=0.5\textwidth]{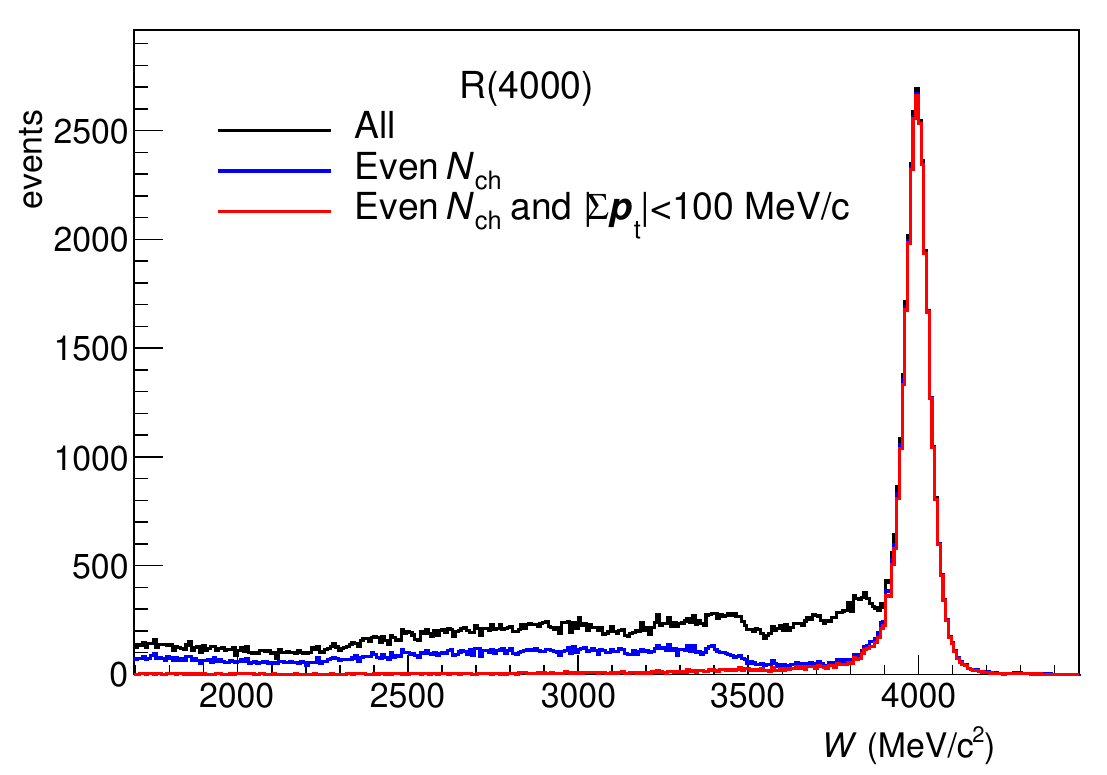}\includegraphics[width=0.5\textwidth]{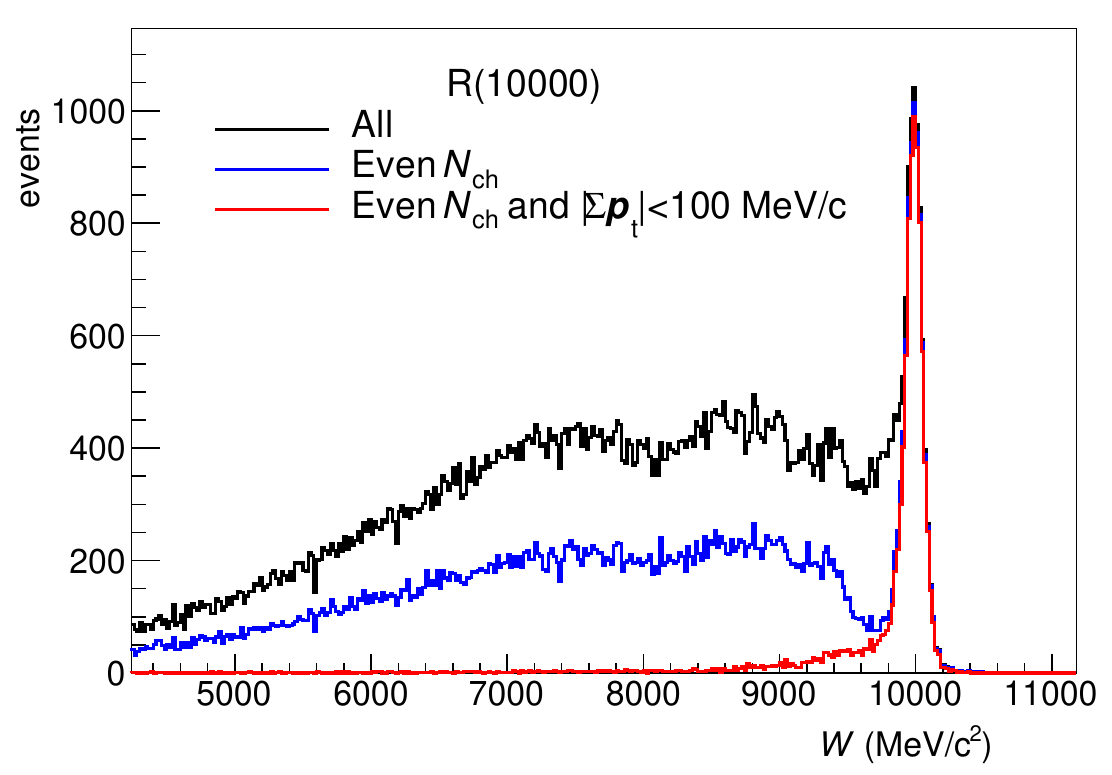}
\caption{Distributions on the invariant masses in the detector for resonances with $M$ = 4 and 10 GeV/$c^2$. Black curves -- all events, blue -- events with an even number of charged particles, red -- with additional cut on the total transverse momentum. Hadronic background is not shown here.}
\label{w-3curves}
\end{figure*}

\begin{figure*}[htbp]
\centering
\includegraphics[width=0.5\textwidth]{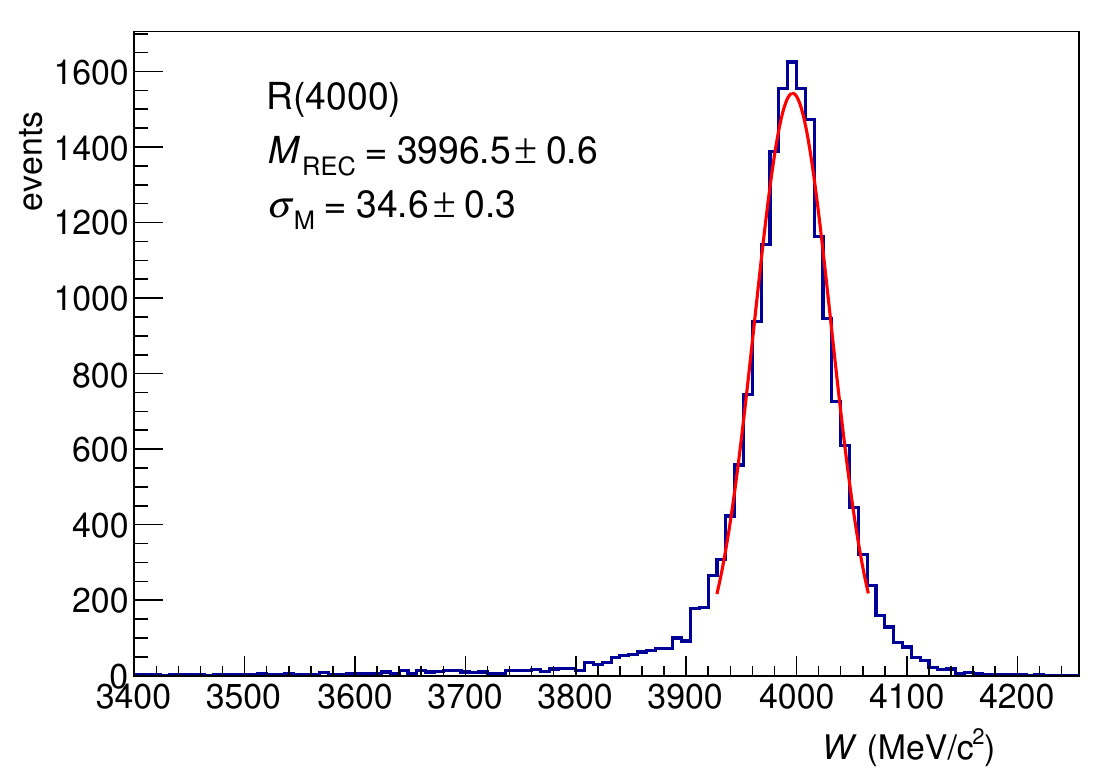}\includegraphics[width=0.5\textwidth]{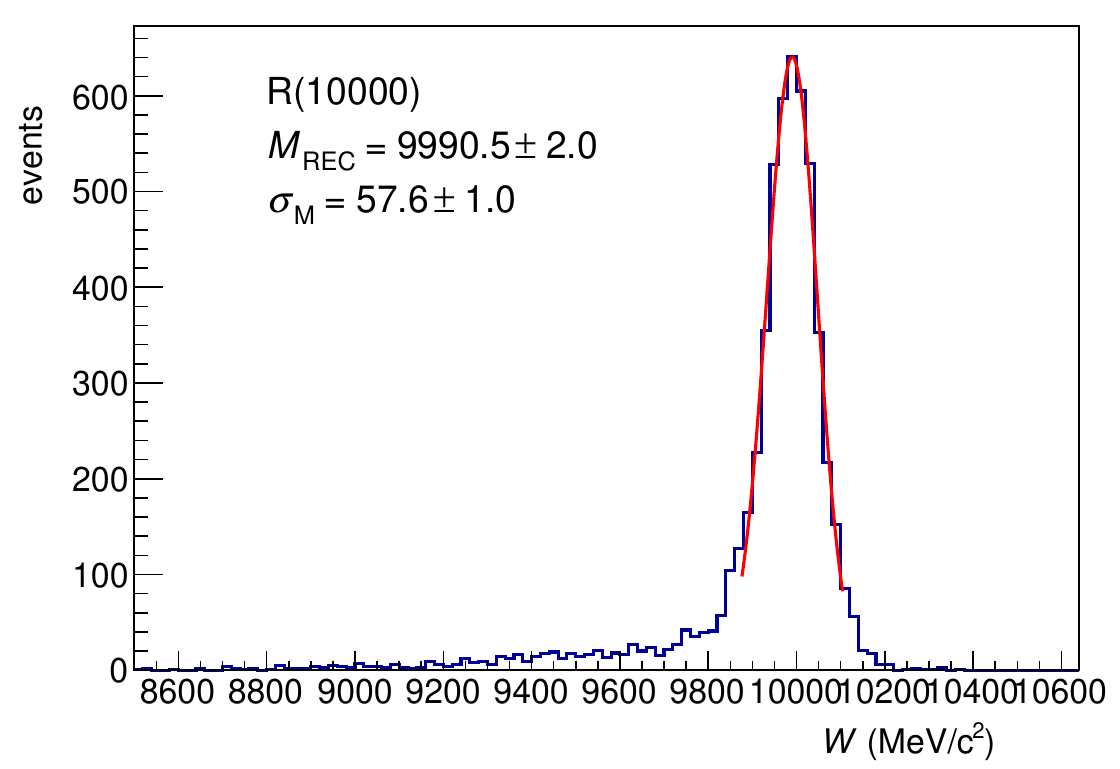}
\caption{Distribution of resonances with $M$ = 4 and 10 GeV/$c^2$ on invariant masses in the detector after the cut on the sum transverse momentum (red curves in Fig.~\ref{w-3curves})   plus cuts 1) and 2) (see the text) which suppress hadronic background. Hadronic background is not shown here.}
\label{w-peaks}
\end{figure*}

     The distributions of the invariant masses in the detector are shown in Fig.~\ref{w-3curves}. There are three distribution: all events, with an even number of charged particles, and with the cut  $|\Sigma p_t|<100$ MeV/c. The last condition leaves only events at the peak of the resonance. After adding constraints 1) and 2) to Fig.~\ref{w-3curves}, we obtain final distributions of invariant masses for resonances shown in Fig.~\ref{w-peaks}.

\begin{figure}[htbp]
\centering
\includegraphics[width=0.5\textwidth]{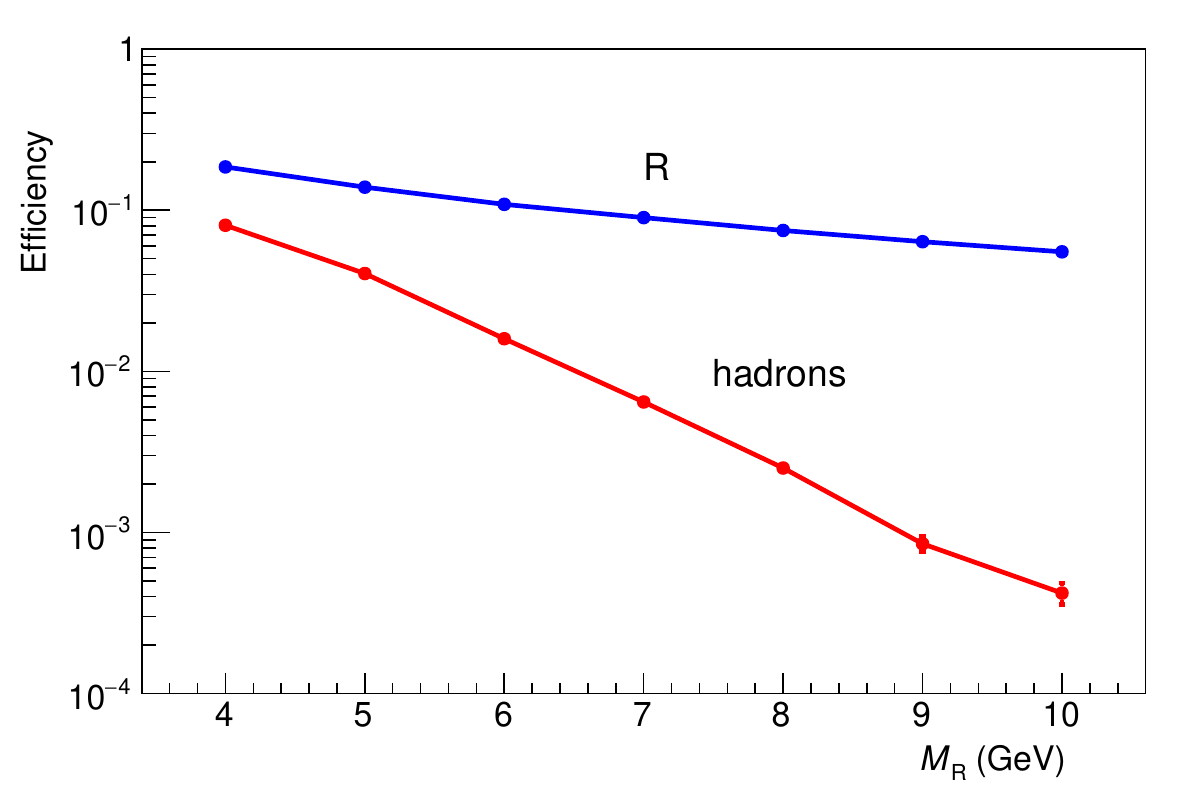}
\caption{Efficiencies for resonances and hadrons after applying all selection criteria.}
\label{eff}
\end{figure}

     The efficiencies for resonances $\epsilon_R$ and hadronic background $\epsilon_h$ after applying  all the selection criteria are presented in Fig.~\ref{eff}. The efficiency for resonances varies from 18\% to 5\% for $M_R=$ 4\,--\,10 \gevcc. The efficiency for the hadron background is lower than for resonances by a factor of 2.5 times  at  $W=4$ \gevcc, and a factor of 125 at 10 \gevcc. This is due to the fact that at higher energies the hadronic background is directed more forward and  differs more from isotropic (at $J$ = 0) decays of resonances.  Such a behavior was expected but was not quantified. This result shows the possibility of studying $C$-even resonances in \GG\ collisions in the energy region of 10 GeV, where the ratio of the non-resonant hadronic cross section to the bottomonium resonance cross sections is two orders of magnitude greater than in the charmonium energy region of $W=$ 3\,--\,4 \gevcc. The above analysis assumed a minimum detector angle of 0.15 rad. We repeated the same analysis for the angle 0.3 rad. For $W=10$ \gevcc the decrease in efficiency is about 15\% and negligible for $W=4$ \gevcc. The difference is so small because we selected events with a large sum of transverse momenta to suppress the hadronic background.

\begin{figure}[htbp]
\centering
\includegraphics[width=0.5\textwidth]{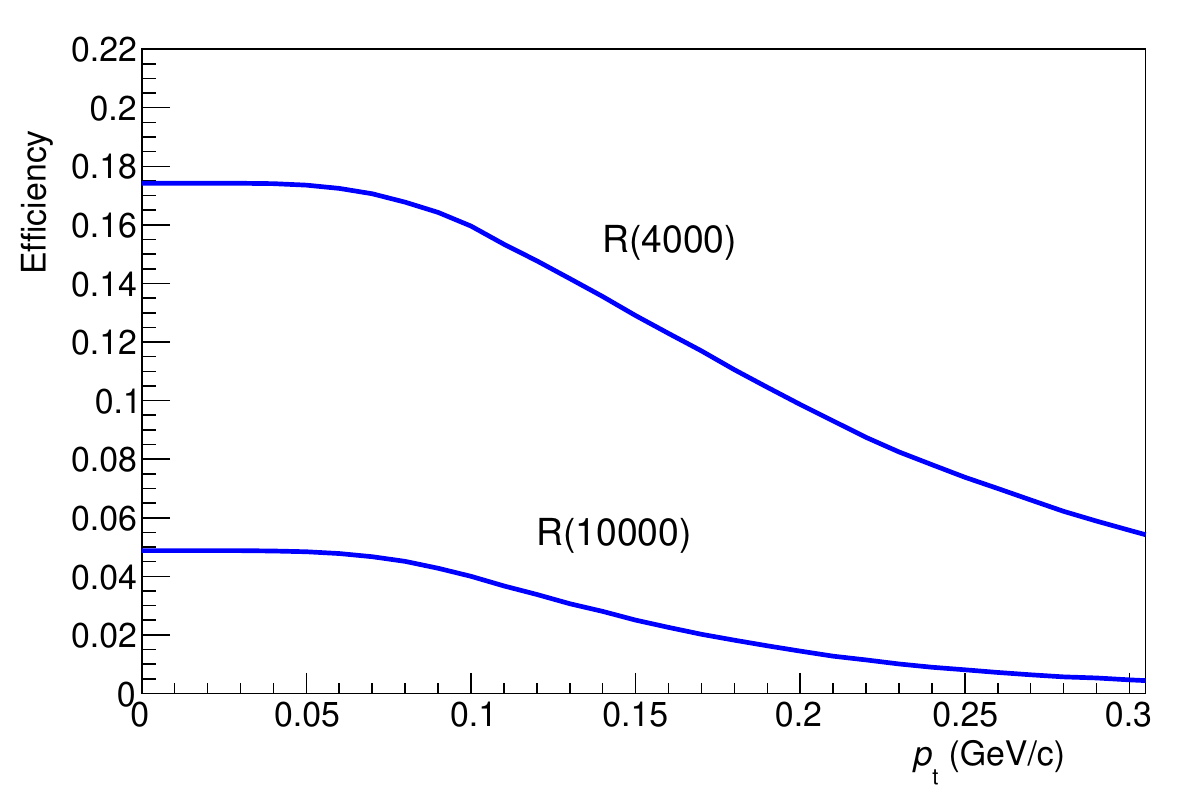}
\caption{Efficiencies for resonances after all selection criteria (as in Fig.\ref{eff}) with an additional cut on the minimum  $p_t$ of particles in the detector (if it will be needed for suppression of low $p_t$ QED backgrounds, see the text).}
\label{ptmin}
\end{figure}

     Fig.~\ref{ptmin} shows how efficiency decreases when an additional cut is applied on the minimum $p_t$ of charged particles in the detector. This information is useful when considering QED backgrounds with a small $p_t$. It comes mainly from low-energy $\GG \to \epem$ process. For the present analysis, the effective $p_{t, \rm{min}} \approx 50$ MeV/$c$, as it is seen in Fig.~\ref{ptmin}. The background $e^+$ and $e^-$ overlap with the events under study with a probability of about 2\% for the collider parameters corresponding to Fig.~\ref{pol-np}. Such low $p_t$ tracks, identified as $e^+/e^-$, can simply be ignored in event analysis because the probability of such particles in the decay products of resonances is very small. The probability of imposing a hadron background on the effect is about 1.5\%.

\begin{figure}[htbp]
\centering
\includegraphics[width=0.5\textwidth]{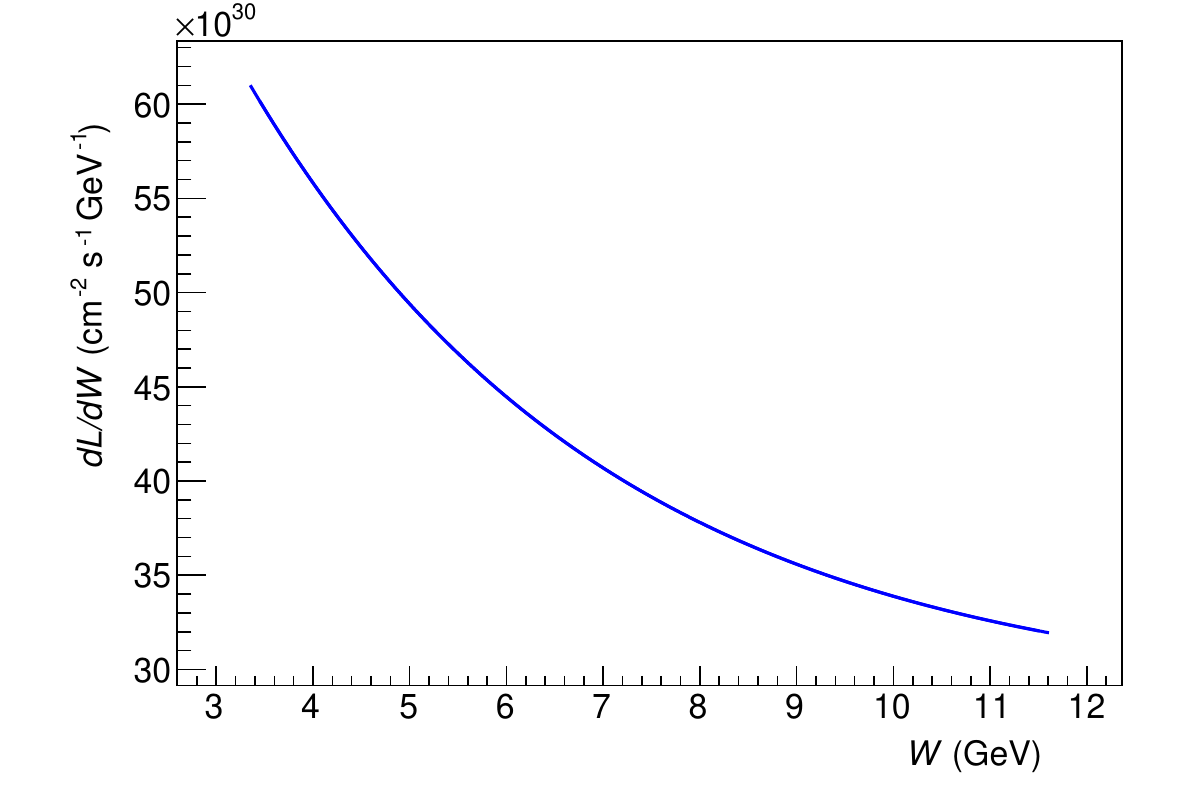}
\caption{The differential luminosity $dL/dW$ at the high energy peak of luminosity spectra (at the photon collider under consideration) as a function of $W_{\mathrm peak}$ which is varied by the electron energy.}
\label{dldw}
\end{figure}

\begin{figure}[htbp]
\centering
\includegraphics[width=0.5\textwidth]{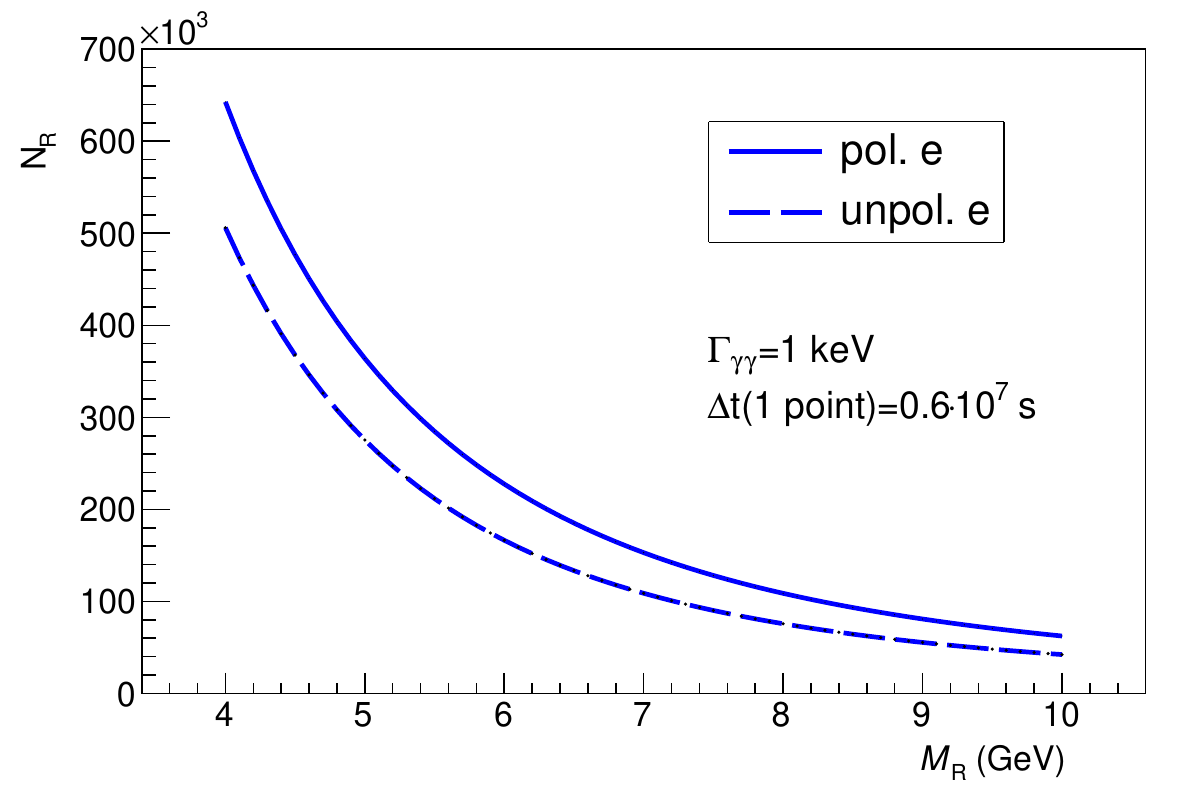}
\caption{The number of produced resonances (no cuts) with $\GGG= 1$ keV for the running time at one energy point equal to 1/5 of the year.}
\label{nres}
\end{figure}

     Fig.\ref{dldw} shows the differential luminosity $dL/dW$ of the \GG\ collider under consideration at the high energy peak of luminosity spectra as a function of $W$ (which is varied by the electron energy).  The number of produced resonances (no cuts) with $\GGG= 1$ keV for the running time at one energy point equal to 1/5 of the year is shown in Fig.~\ref{nres}.

\begin{figure}[htbp]
\centering
\includegraphics[width=0.5\textwidth]{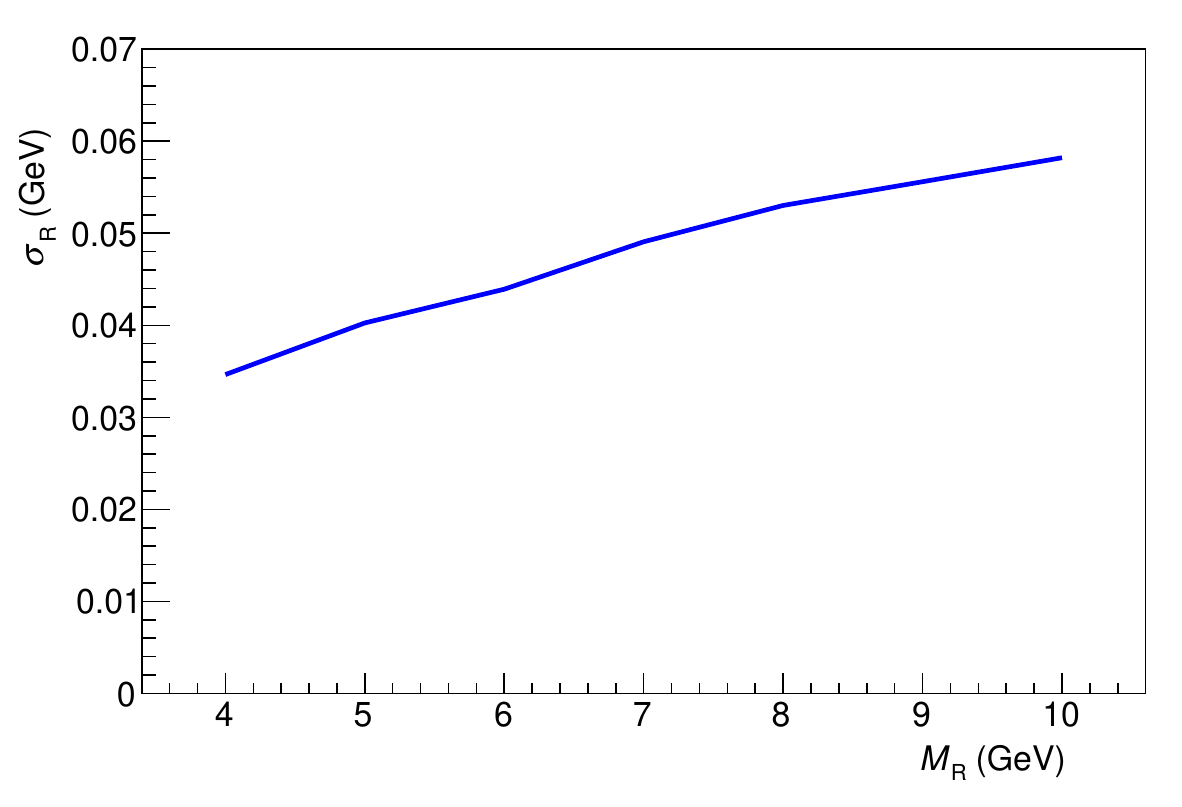}
\caption{The mass resolution for resonances.}
\label{mres}
\end{figure}

\begin{figure}[htbp]
\centering
\vspace{1cm}
\includegraphics[width=0.5\textwidth]{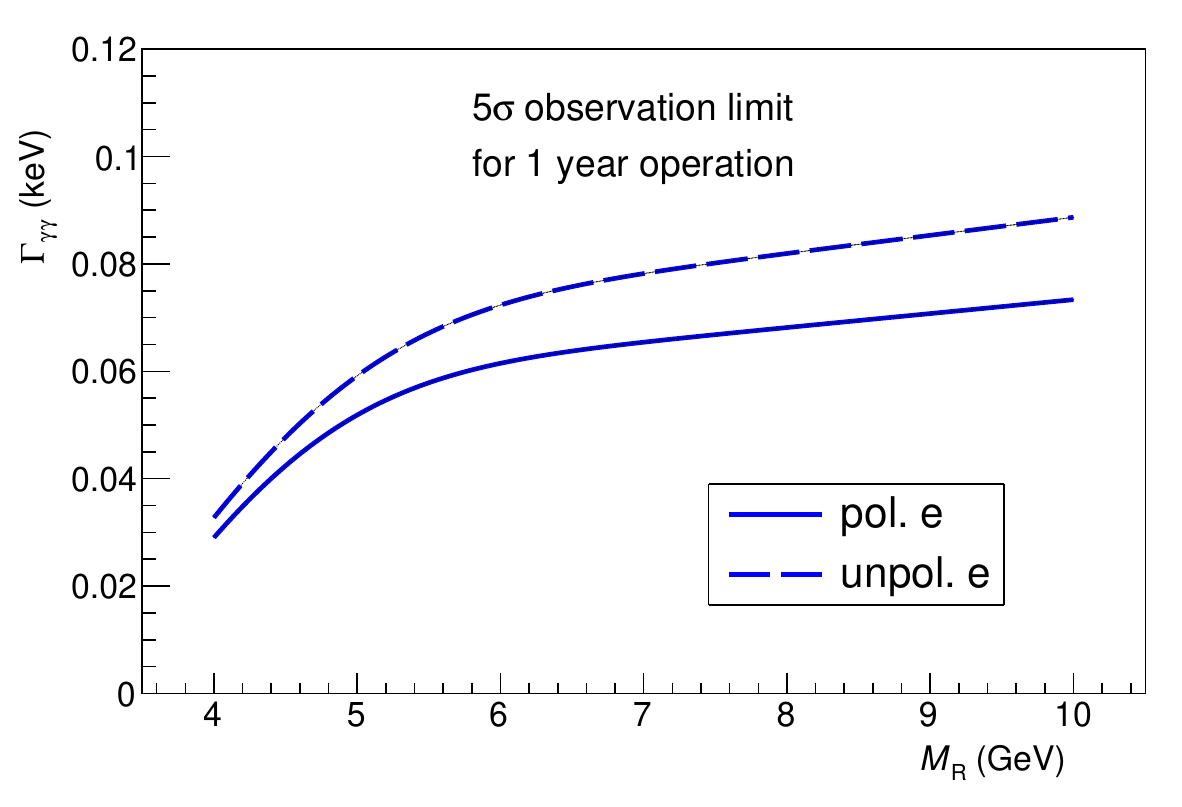}
\caption{The minimum values of $\Gamma_{\GG}$ for detecting resonances at the 5 sigma level for 1/5 year operation on the energy of the resonance.}
\label{5sig}
\end{figure}

     The mass resolution of reconstructed resonances is given in Fig.~\ref{mres}. It is $\sigma_{M_R}\approx$ 35\,--\,55 \mevcc for $W=$ 4\,--\,10 \gevcc for the chosen detector parameters.  The minimum values of $\Gamma_{\GG}$ for detecting resonances at the 5$\sigma$ confidence level in 1/5 year operation on the energy of the resonance is given Fig.~\ref{5sig}. In this case, about 5 energy points (one year) cover  the entire region of invariant masses. It was assumed in calculations that $\sigma_{\GG \to \rm{hadr}} \approx 350 \nb$  in mass region $W=$ 4\,--\,10 \gevcc.
     The ratio of the resonance peak height to the non-resonant background
  % \vspace{-0.5cm}
  \be
        R=\frac{dN_R/dW}{dN_h/dW}=
     \frac{4\pi^2(2J+1)\Gamma_{\GG}(1+\lambda_{\gamma}\tilde{\lambda}_{\gamma})\epsilon_R}{\sqrt{2\pi}M_R^2\sigma_{M_R} \sigma_h \epsilon_h}.
       \ee
  %    \vspace{-0.3cm}
     For the lightest  $C$-even charmonium $\eta_c(2984)$ with $\Gamma_{\GG}\approx 5$ keV and the lightest bottonium $\eta_b(9398)$ with $\Gamma_{\GG}\sim 0.5$ keV the values of $R$ are approximately 1.4 and 0.4, respectively. The $\eta_b(9398)$ meson has not yet been observed in the $\GG$ mode; at the photon collider it can be observed at the >5$\sigma$ level in one day of operation.

 \section{Conclusion}
     Our analysis showed that the hadron background in the \bbbar energy region ($W \sim 10$ \gevcc) can be suppressed by more than two orders of magnitude, which makes it possible to study $C$-even resonances at the \GG\ collider with masses up to 12 \gevcc by detecting  all final particles (all hadronic decay modes together).  As can be seen in Fig.\ref{gammavsm}, the region $W=$ 3\,--\,12 \gevcc is populated by many resonance states of various nature, which can be studied at the photon collider on the base of the European XFEL linac (or at any other photon collider at these energies).

\section*{Acknowledgments}

  This work was supported by RFBR-DFG Grant No 20-52-12056.

%\clearpage

%

\vspace{-0.5cm}
\vspace{-0.5cm}



\end{document}